# Lattice Design in High-Energy Particle Accelerators

*B. J. Holzer*
CERN, Geneva, Switzerland

**Abstract**
This lecture gives an introduction into the design of high-energy storage ring lattices. Applying the formalism that has been established in transverse beam optics, the basic principles of the development of a magnet lattice are explained and the characteristics of the resulting magnet structure are discussed. The periodic assembly of a storage ring cell with its boundary conditions concerning stability and scaling of the beam optics parameters is addressed as well as special lattice insertions such as drifts, mini beta sections, dispersion suppressors, etc. In addition to the exact calculations that are indispensable for a rigorous treatment of the matter, scaling rules are shown and simple rules of thumb are included that enable the lattice designer to do the first estimates and get the basic numbers 'on the back of an envelope'.

## 1 Introduction

Without doubt the highlight of the present year in high-energy physics is the discovery of the Higgs particle at CERN and as a consequence, and nice side effect for those involved, the Nobel price in physics begin awarded to Professors Higgs and Englert. Within a remarkably short period, namely during the LHC run 1 in 2009–2012, the high-energy physics detectors installed at the LHC could collect a sufficient amount of data to prove the existence of a new particle, identified as a Higgs boson. A state-of-the-art picture of a very clear example of such an 'event', detected by the ATLAS group at CERN, is shown in Fig. 1 where the four decay products of the Higgs, i.e. two muons and two electrons with a centre of mass energy of $E_{cm}$ = 126 GeV, are clearly visible.

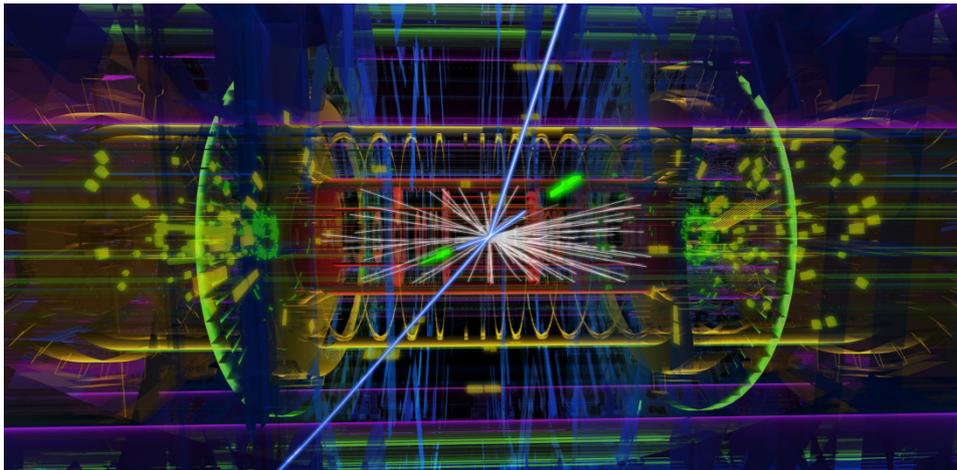

**Fig. 1:** ATLAS event display of a Higgs candidate decaying into two electrons and two muons (courtesy of the ATLAS collaboration).

At present (status year 2013) about 1400 Higgs events could be identified in the detectors CMS and ATLAS with a luminosity of the LHC storage ring of up to $L = 7 \times 10^{33}$ cm$^{-2}$ s$^{-1}$, which is close to the design value of the machine. We will explain later (Section 5) how the luminosity of a particle collider can be optimized. For the time being we simply state that the event rate at a collision point of two particle beams depends on the physics cross-section of the process, $\Sigma_{react}$, and a number that describes the performance of the storage ring, i.e. the luminosity

$$R = L * \Sigma_{react} \quad (1)$$

A prominent example that we already mentioned, the cross-section of the Higgs for different production processes is shown in Fig. 2. As a rule of thumb it lies in the range of $\Sigma_{Higgs} = 1$ pb. The luminosity that could be integrated in the LHC over the operation period of run 1 is plotted on the right-hand side of the figure: approximately $L = 25$ fb$^{-1}$ could be delivered to each experiment.

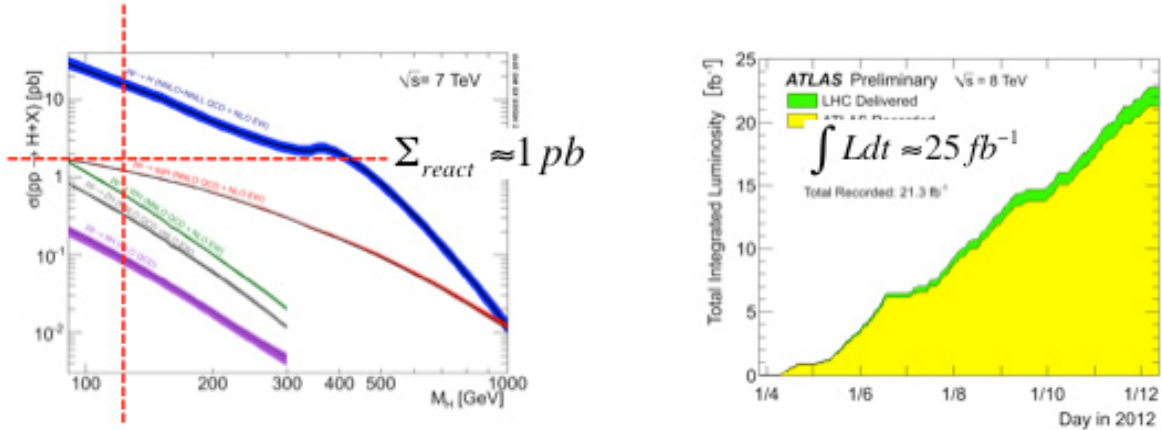

**Fig. 2:** Higgs cross-section and integrated luminosity of the LHC collider in run 1

Just as in the case of the LHC and the search for the Higgs boson the study of nuclear and high-energy physics has always been the driving force for the development and optimization of high-energy particle beams; while the first experiments in that field were performed using 'beams' from natural radioactive particle sources such as α or β emitters, it soon became clear that in order to obtain higher-energy particles, special machines, particle accelerators, had to be developed (see e.g. [1]).

The basis for the design of a storage ring and its magnet lattice, which is the topic of this paper, was laid in 1952, when Courant, Livingston and Snyder developed the theory of the so-called strong focusing accelerators (or alternating gradient machines) [2].

Lattice design in the context we will describe it here is the design and optimization of the principle elements, the lattice cells, of a (circular) accelerator, and it includes the dedicated variation of the lattice elements (as for example position and strength of the magnets in the machine) to obtain well-defined and predictable parameters of the stored particle beam.

It is therefore closely related to the theory of linear beam optics that has been described in several text books and lecture notes (e.g. CAS introductory school [3, 4]) and accordingly we will only repeat briefly the main concepts to be able to apply them in a decent way: the design of a machine lattice is in that sense an application of the linear beam optics and as it has a high practical relevance we will try in this contribution to dilute the formalism a bit by introducing here and there lattice examples of present day storage rings and plots from optics calculations of real machines.

As a first example the beam optics in a part of a lattice structure is shown in Fig. 3. The plot shows the most important beam parameters in a typical high-energy storage ring. In the upper part the $\beta$ function in the horizontal $\beta_x$ and vertical $\beta_y$ plane is plotted (or rather its square root which is proportional to the beam size), in the middle part the position of the lattice elements is shown and the lower part shows the so-called dispersion function in $x$ and $y$. We will discuss these parameters in some detail in the following sections of this paper.

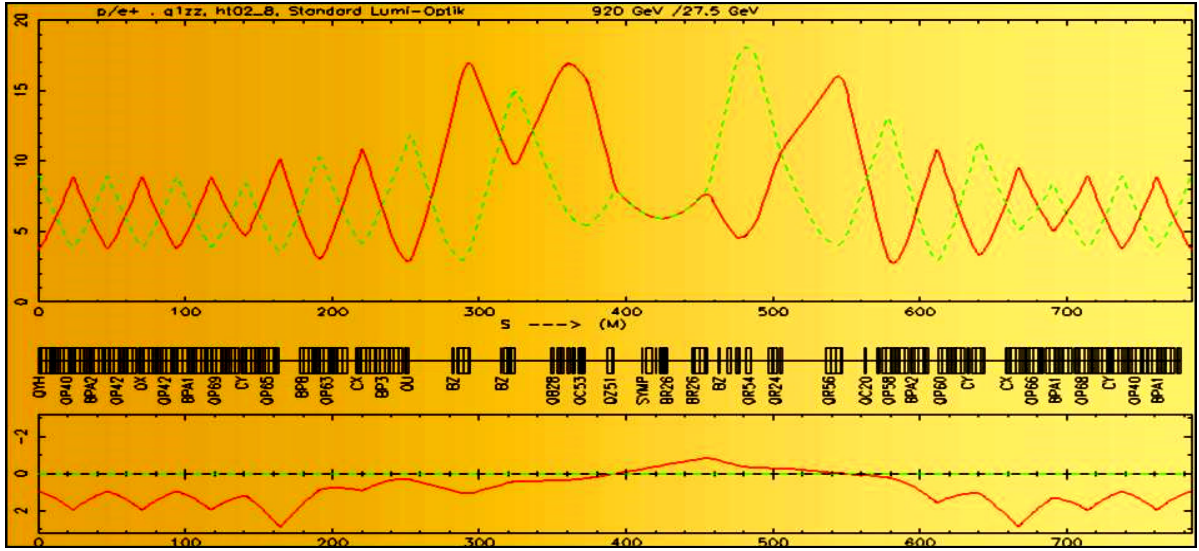

**Fig. 3:** Lattice and beam optics in a part of a typical high-energy accelerator. The curves in the upper part refer to the square root of the $\beta$ function, the lower part shows the dispersion.

## 2    Geometry of the ring

The first step in the layout of a storage ring is the fixing of the geometry or, which is closely related to that, the determination of the momentum of the particles that will be stored in it.

For the bending force as well as for the focusing of the particle beam, magnetic fields are applied in a circular accelerator. In principle, electrostatic fields would be possible as well but at high momenta (i.e. if the particle's velocity is close to the speed of light) magnetic fields are much more efficient. The force acting on the particles, the Lorentz force, is given by

$$F = q(E + v \times B).  \qquad (2)$$

Now from Eq. (2) it is evident that each magnetic field that is applied to influence the particle beam gets a nice amplification by the value of the speed of the particles and as we are talking here mainly about high-energy particle beams and thus $v \approx c$, we conclude that in this case magnetic fields are much more efficient than electrical ones.

Assuming a magnetic $B$ field of $B = 1$ T acting perpendicular to the speed of the particles we obtain

$$\mathbf{v} \times \mathbf{B} = v \cdot B \approx 3 \times 10^8 \, \mathrm{m/s} \cdot \mathrm{Vs/m}^2$$
$$\approx 300 \, \mathrm{MV/m}.$$

A magnetic $B$ field of 1 Tesla, which is easy to create and maintain, corresponds to an electrical field of $E = 300$ MV/m, which is far beyond any technical limit ($E < 1$MV/m).

Neglecting the **E**-field for the time being and assuming a constant transverse magnetic dipole field **B**, the particle will see a constant deflecting force and the trajectory will be a part of a circle. In other words, the condition for a circular orbit is that the Lorentz force is equal to the centrifugal force:

$$\frac{\gamma m_0 v^2}{\rho} = evB.$$

Dividing by the velocity $v$ we obtain a relation between the magnetic field and the momentum of the particle

$$e \cdot B = \frac{mv}{\rho} = \frac{p}{\rho}.$$

The term $B\rho$ is called the beam rigidity,

$$B\rho = \frac{p}{q}$$

and connects the magnetic dipole field needed for a circular orbit of radius $\rho$ to the particle's momentum and charge. (Note that here we often refer to protons or electrons and the charge is just the elementary charge $e$.) Given an ideal circular orbit, for each segment of the path we obtain the relation

$$\alpha = \frac{ds}{\rho} = \frac{B\,ds}{B\rho}$$

and integrating along the path $ds$ in all dipole magnets in the ring we require

$$\alpha = \frac{\int B\,ds}{B\rho} = 2\pi \rightarrow \int B\,ds = 2\pi \frac{p}{q}, \tag{3}$$

where we have replaced the elementary charge $e$ by the more general expression of the particle charge $q$. As in any 'circular' accelerator the angle swept in one turn for the design particle is $2\pi$, Eq. (3) tells us that the integral of all bending magnets in the ring has to be $2\pi$ times the momentum of the beam. If the path length inside the dipole magnet differs not too much from the length of the magnet itself, the integral in Eq. (3) can be approximated by $\int B\,dl$ where $dl$ refers to the magnet length.

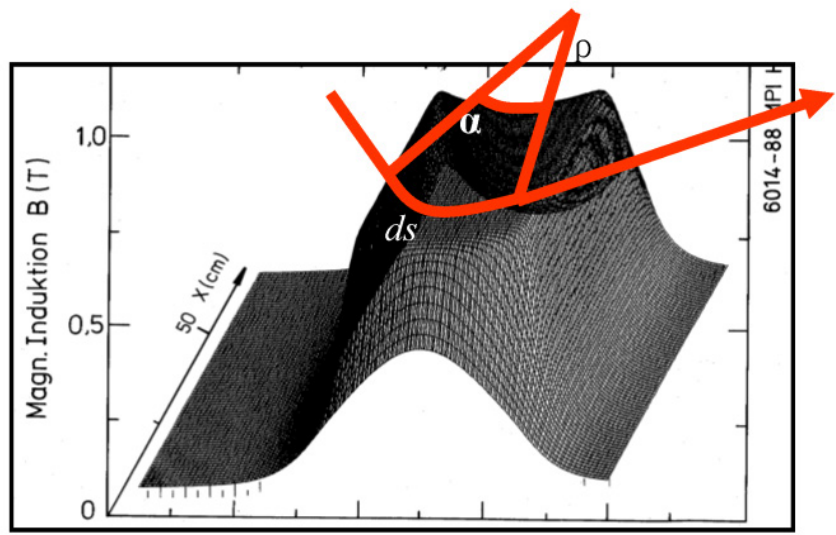

**Fig. 4:** Map of the magnetic $B$ field in a storage ring dipole magnet

The field of a typical bending magnet as used in a storage ring is shown in Fig. 4. On the vertical scale the magnetic induction $B$ is shown in Tesla, measured between the two pole faces of the dipole magnet. The plateau of constant field is beautifully seen inside the magnet as well as the decreasing edge fields before and after the magnet.

For the lattice designer the integrated *B* field along the particles design orbit (roughly sketched in the figure) is the most important parameter, as it is the value that enters Eq. (3) and defines the field strength and the number of such magnets that are needed for a full circle.

Figure 5 shows a photograph of a small storage ring [5] where only eight dipole magnets are used to define the design orbit. The magnets are powered symmetrically and therefore each magnet provides a bending angle of exactly $\alpha = 45°$ of the beam. The field strength in this machine is of the order of $B = 1$ Tesla.

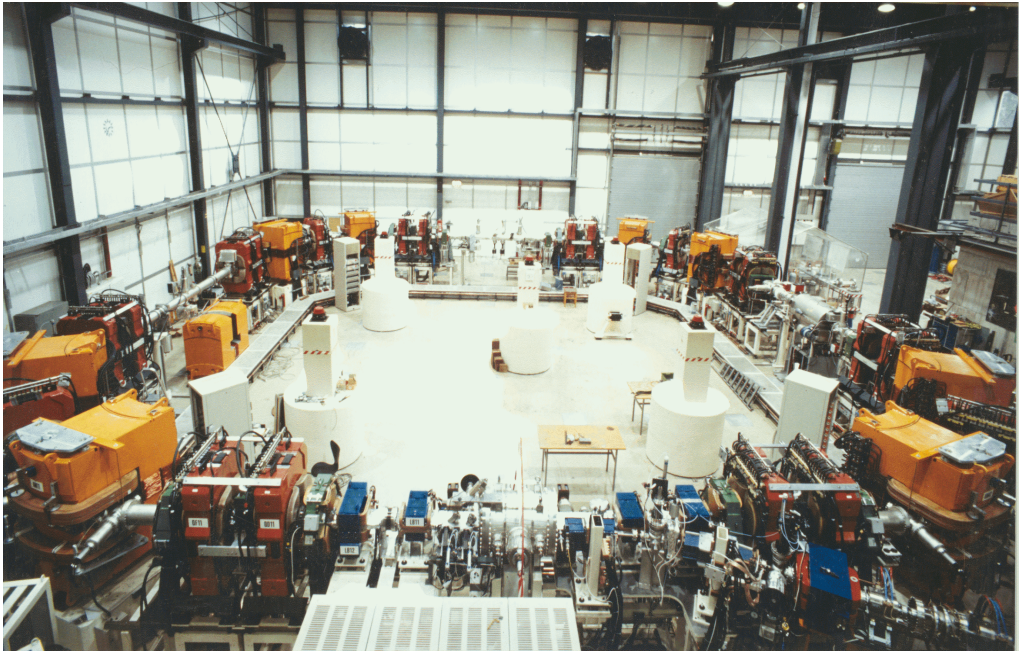

**Fig. 5:** Heavy ion storage ring TSR of the Max-Planck Institute in Heidelberg

In general, unlike in the example in Fig. 5, for a high-energy storage ring or synchrotron a large number of bending magnets is needed to determine the design orbit, and very high magnetic fields.

As most prominent example the LHC storage ring is presented here (see Fig. 6): it accelerates and stores proton beams of an energy of up to 7 TeV and collides them at four interaction points. In the LHC, 1232 dipole magnets are used with a length of $l = 15$ m each, to guide the beam on an orbit of nearly 27 km circumference.

At an energy of 7 TeV the particles are ultra relativistic and particle energy and momentum are proportional to each other

$$E \approx pc,$$

thus we can calculate the momentum as a function of the *B* field:

$$\int B \, dl \approx NlB = 2\pi p / e. \tag{4}$$

Using the formula (4), for the magnetic field we obtain

$$B \approx \frac{2\pi \; 7 \times 10^{12} \text{ eV}}{1232 \cdot 15 \text{ m } 3 \times 10^8 \text{ m/s } e} = 8.3 \text{ Tesla}.$$

It is immediately clear that the machine had to be built with superconducting magnets to achieve the design energy of 7 TeV.

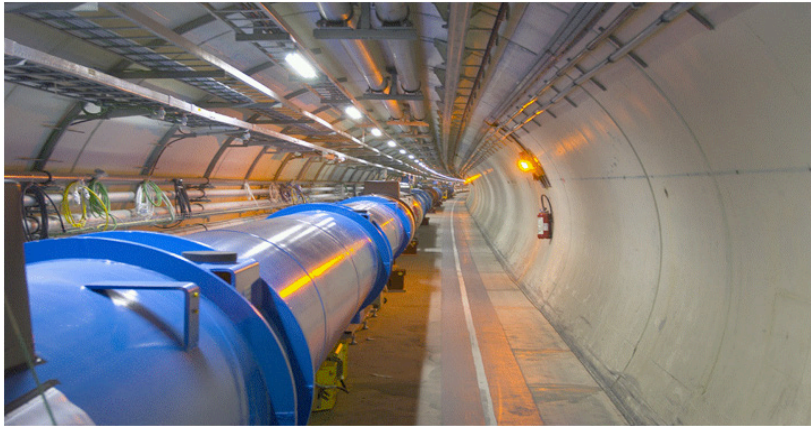

**Fig. 6**: LHC storage ring: 1232 super conducting magnets are needed to bend the protons on a circular path of 27 km length.

## 3  Equation of motion and matrix formalism

If the geometry and specification of the arc is determined and the layout of the bending magnets is done, the next step is to worry about the focusing properties of the machine. In general we have to keep more than $10^{12}$ particles in the machine, distributed in a certain number of bunches. And these particles have to be focused to keep their trajectories close to the design orbit.

As we have heard in the linear beam optics lecture, gradient fields are used to do the job. They generate a magnetic field that is increasing linear as a function of the distance to the magnet centre:

$$B_y = -gx, \qquad B_x = -gy.$$

Here $x$ and $y$ refer to the horizontal and vertical plane and the parameter $g$ is called the gradient of the magnetic field. It is custom to normalize the magnetic fields to the momentum of the particles. In the case of the dipole fields we obtain from Eq. (3)

$$\alpha = \frac{\int B \, ds}{B\rho} = \frac{L_{\text{eff}}}{\rho}.$$

$L_{\text{eff}}$ is the so-called effective length of the magnet and to be very precise the integral should be taken along the particle trajectory, which in a dipole magnet is *not* a straight line. The term $1/\rho$ is the normalized bending strength of the dipole. In the same way the field of the quadrupole lenses is normalized to the beam rigidity $B \cdot \rho$: the strength $k$ is defined by

$$k = \frac{g}{B \cdot \rho}$$

and the focal length of the quadrupole is given by

$$f = \frac{1}{k \cdot l_q}.$$

Under the influence of the focusing properties of the quadrupole and dipole fields in the ring the particle trajectories are described by a differential equation. In the lecture about linear beam optics this equation is derived in full beauty, so here we just state that it is given by the expression

$$x'' + Kx = 0. \tag{5}$$

Here $x$ describes the horizontal coordinate of the particle with respect to the design orbit, the derivative is, as usual in linear beam optics, taken with respect to the orbit coordinate $s$, $x' = dx/ds$ and the parameter $K$ combines the focusing strength $k$ of the quadrupole and the weak focusing term $1/\rho^2$ of the dipole field (note that a negative value of $k$ means a horizontal focusing magnet):

$$K = -k + \frac{1}{\rho^2}.$$

In the vertical plane in general the term $1/\rho^2$ is missing, as in most accelerators (but not in all) the design orbit is in the horizontal plane and no vertical bending strength is present. So we get in the vertical plane

$$K = +k.$$

We want to point out that when starting to design a magnet lattice we would like to make as many simplifications as possible in the beginning. Clearly the exact solution of the particle motion has to be calculated in full detail and if the beam optics is optimized on a linear basis, higher-order multipole fields and their effect on the beam have to be taken into account. But doing the very first steps we can make life a little bit easier and ignore terms that are small enough to be neglected. In many cases for example the weak focusing term $1/\rho^2$ can be neglected to get a rough estimate which makes the formula much shorter and symmetric in the horizontal and vertical plane. Referring again to the example of the LHC storage ring, the basic parameters of the machine are summarized in Table 1.

**Table 1:** Basic parameters of the LHC storage ring

| | |
|---|---|
| Circumference $C_0$ | 26,659 m |
| Bending radius $\rho$ | 2804 m |
| Quadrupole gradient $G$ | 223 T/m |
| Particle momentum $p$ | 7000 GeV/$c$ |
| Weak focusing term $1/\rho^2$ | $1.27 \times 10^{-7}$ 1/m$^2$ |
| Focusing strength $k$ | $8.6 \times 10^{-3}$ 1/m$^2$ |

The weak focusing contribution $1/\rho^2 = 1.27 \times 10^{-7}$/m is indeed much smaller than the quadrupole strength $k$. For first estimates in the lattice of large accelerators therefore it can in general be neglected.

## 3.1 Single-particle trajectories

The differential equation, Eq. (5), describes the transverse motion of the particle with respect to the design orbit. In linear approximation it can be solved and the solutions for the horizontal and vertical planes are independent of each other.

If the focusing parameter $K$ is constant, that means if we refer to the situation to a place inside a magnet where the field is constant along the orbit, the general solution for the position and angle of the trajectory can be derived as a function of the initial conditions $x_0$ and $x'_0$. In the case of a focusing lens we obtain

$$x(s) = x_0 \cdot \cos(\sqrt{K} \cdot s) + \frac{x'_0}{\sqrt{K}} \sin(\sqrt{K} \cdot s),$$
$$x'(s) = -x_0 \sqrt{K} \sin(\sqrt{K} \cdot s) + x'_0 \cos(\sqrt{K} \cdot s).$$

Or, written in a more convenient matrix form,

$$\begin{pmatrix} x \\ x' \end{pmatrix}_{s1} = M_{\text{foc}} \begin{pmatrix} x \\ x' \end{pmatrix}_{s0}.$$

The matrix *M* depends on the properties of the magnet and for a number of typical lattice elements we get the following expressions:

$$M_{QF} = \begin{pmatrix} \cos(\sqrt{K} \cdot s) & \frac{1}{\sqrt{K}} \sin(\sqrt{K} \cdot s) \\ -\sqrt{K} \cdot \sin(\sqrt{K} \cdot s) & \cos(\sqrt{K} \cdot s) \end{pmatrix} \quad (6a)$$

for a focusing quadrupole

$$M_{QD} = \begin{pmatrix} \cosh(\sqrt{K} \cdot s) & \frac{1}{\sqrt{K}} \sinh(\sqrt{K} \cdot s) \\ \sqrt{K} \cdot \sinh(\sqrt{K} \cdot s) & \cosh(\sqrt{K} \cdot s) \end{pmatrix} \quad (6b)$$

for a defocusing quadrupole, and

$$M_{drift} = \begin{pmatrix} 1 & l \\ 0 & 1 \end{pmatrix} \quad (6c)$$

for a drift space.

### 3.2 The Twiss parameters α, β, γ

In the case of periodic conditions in the accelerator there is another way to describe the particle trajectories which is in many cases more convenient than the above-mentioned formalism which is valid within a single element.

It is important to note that in a circular accelerator the focusing elements are necessarily periodic in the orbit coordinate *s* after one revolution. Even more, storage ring lattices have in most cases an inner periodicity: they often are built, at least partly, of sequences where identical magnetic cells, the lattice cells, are repeated several times in the ring and lead to periodically repeated focusing properties.

In this case the transfer matrix from the beginning of such a structure to the end is expressed as a function of the periodic parameters *α, β, γ and the phase φ*:

$$M(s) = \begin{pmatrix} \cos\varphi + \alpha_s \sin\varphi & \beta_s \sin\varphi \\ -\gamma_s \sin\varphi & \cos\varphi - \alpha_s \sin\varphi \end{pmatrix}. \quad (7)$$

The parameters *α* and *γ* are related to the *β* function by the equations

$$\alpha(s) = -\frac{1}{2}\beta'(s) \quad \text{and} \quad \gamma(s) = -\frac{1+\alpha^2(s)}{\beta(s)}.$$

The matrix is clearly a function of the position *s,* as the parameters *α, β, γ* depend on *s*. The variable *φ* is called the phase advance of the trajectory and is given by

$$\varphi_{12} = \int_{s1}^{s2} \frac{d\tilde{s}}{\beta(\tilde{s})}.$$

In such a periodic lattice and referring to one plane at a time, for stability of the equation of motion the relation

$$|trace(M)| < 2$$

has to be valid, which sets boundary conditions for the focusing properties of the lattice, as we will see in a moment.

Given that correlation, the solution of the trajectory of a particle can be expressed as a function of these new parameters:

$$x(s) = \sqrt{\varepsilon}\sqrt{\beta(s)} \cdot \cos(\varphi(s) - \varphi_0)$$

$$x'(s) = \frac{-\sqrt{\varepsilon}}{\sqrt{\beta(s)}} \cdot \{\sin(\varphi(s) - \varphi_0) + \alpha(s) \cdot \cos(\varphi(s) - \varphi_0)\}.$$

The position and angle of the transverse oscillation of a particle at a point $s$ is given by the value of the $\beta$ function at that location and $\varepsilon$ and $\varphi_0$ are constants of the particular trajectory.

As a last reminder of the linear beam optics we state that the Twiss parameters at a position $s$ in the lattice are defined by the focusing properties of the complete storage ring. They are transformed through the lattice from one point to another by the matrix elements of the corresponding magnets. Without proof we state that if the matrix $M$ is given by

$$M(s_1, s_2) = \begin{pmatrix} C & S \\ C' & S' \end{pmatrix}$$

the transformation rule from point $s_1$ to $s_2$ in the lattice is obtained via

$$\begin{pmatrix} \beta \\ \alpha \\ \gamma \end{pmatrix}_{s2} = \begin{pmatrix} C^2 & -2SC & S^2 \\ -CC' & SC'+CS' & -SS' \\ C'^2 & -2S'C' & S'^2 \end{pmatrix} \cdot \begin{pmatrix} \beta \\ \alpha \\ \gamma \end{pmatrix}_{s1}. \tag{8}$$

The terms $C$, $S$, etc., correspond to the focusing properties of the matrix. In the case of one single element, for example, they are just the expressions given in Eq. (6). For a longer structure it is the resulting product matrix elements that will be used.

## 4   Lattice design

To illuminate these facts we start with the investigation of the easiest situation: a simple drift space in a lattice.

As shown in Eq. (5) the matrix for a drift space is given by

$$M_{\text{drift}} = \begin{pmatrix} C & S \\ C' & S' \end{pmatrix} = \begin{pmatrix} 1 & l \\ 0 & 1 \end{pmatrix}. \tag{9}$$

Starting with position $x_0$ and angle $x'_0$ the trajectory after the drift therefore will be

$$\begin{pmatrix} x \\ x' \end{pmatrix}_l = \begin{pmatrix} 1 & l \\ 0 & 1 \end{pmatrix} \cdot \begin{pmatrix} x \\ x' \end{pmatrix}_0$$

or, written explicitly,

$$x(l) = x_0 + l \cdot x'_0,$$
$$x'(l) = x'_0.$$

If the drift is located in a circular accelerator or within a periodic part of a lattice, the Twiss parameters are well defined at the start of the drift and they will be transformed according to Eq. (8) from their initial values $\alpha_0, \beta_0, \gamma_0$ via

$$\begin{pmatrix} \beta \\ \alpha \\ \gamma \end{pmatrix}_l = \begin{pmatrix} 1 & -2l & l^2 \\ 0 & 1 & -l \\ 0 & 0 & 1 \end{pmatrix} \cdot \begin{pmatrix} \beta \\ \alpha \\ \gamma \end{pmatrix}_0.$$

Namely the $\beta$ function in the drift develops as

$$\beta(l) = \beta_0 - 2l\alpha_0 + l^2\gamma_0.$$

Calculating the trace of the matrix Eq. (9), we see however that the condition for stability is not fulfilled:

$$|trace(M_{drift})| = 1 + 1 = 2.$$

So we have already learned that a circular accelerator built exclusively out of drift spaces will not be a stable machine (which is a pity, as it would have been a cheap machine).

*Nota bene*: clearly in any storage ring there will be a large number of drift spaces between the focusing elements. The stability criterion however tells us that the magnetic elements and the drift spaces in between them have to be arranged in a way that the resulting lattice cell describes a stable solution in both the horizontal and vertical plane.

Figure 7 shows the lattice and the beam optics of a typical high-energy storage ring (HERA). In the upper part of the figure the square root of the $\beta$ function is shown in both transverse planes. The broad band in the middle is the lattice: as the ring has a circumference of about 6.3 km the single lattice elements can clearly not be distinguished in the figure. In the lower part the dispersion function is shown.

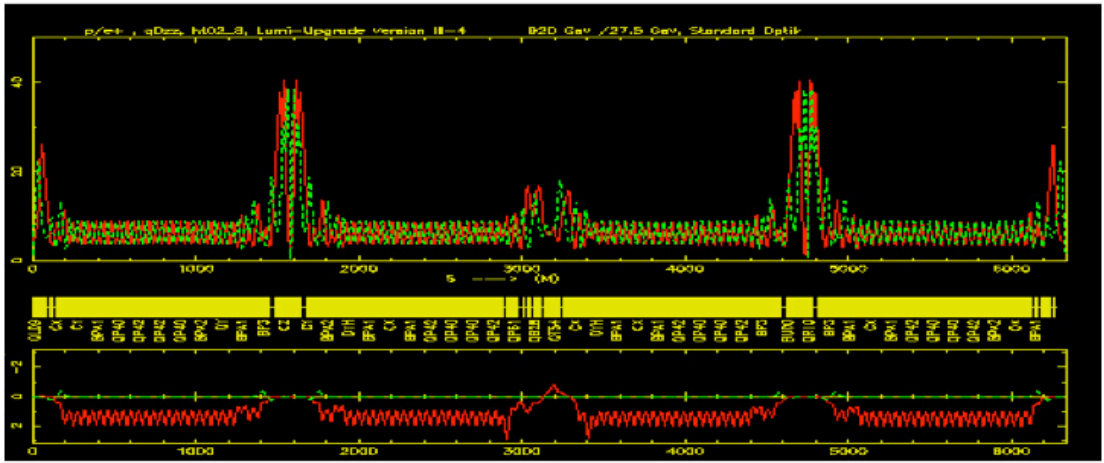

**Fig. 7**: Magnet lattice and luminosity optics of the HERA proton storage ring

On first sight two sections of quite different characteristics can be identified in the accelerator and in fact most if not all high-energy colliders are designed that way.

i) A section where the $\beta$ function shows a regular pattern: these are the arcs where the main bending magnets of the ring are located. They define the geometry of the ring and by the maximum magnetic fields of the bends they also limit the energy (or momentum) of the particle beam. In addition, the main focusing elements of the ring are located in the arcs as well as, in general, quadrupole lenses for tune control and sextupole magnets for the compensation of the chromaticity of the optics.

ii) These regular arcs are connected by straight sections in the ring: long lattice parts where the optics is modified to establish conditions needed for particle injection, to reduce the dispersion function, or where the beam dimensions are reduced to increase the particle collision rate in the case of a collider ring. Beyond that all kind of devices have to be installed in these long sections such as RF cavities, beam diagnostic tools, and even the high-energy particle detectors (if they cannot be avoided in the machine).

Concerning the structure of the arcs it has been shown to be advantageous, to configure them on the basis of small elements, called cells, that repeat each other many times in the ring. One of the most wide-spread lattice cells used for this purpose is the so-called FODO cell which we will discuss in some detail now.

### 4.1 The FODO cell

A magnet structure consisting of focusing and defocusing quadrupole lenses in alternating order with basically nothing in between is called a FODO lattice and the elementary cell a FODO cell. 'Basically nothing' in that context means any element that has only a negligible effect on the focusing properties, as for example drift spaces, RF structures or as we have seen, even bending magnets under certain circumstances. Schematically such a FODO cell is shown in Fig. 8.

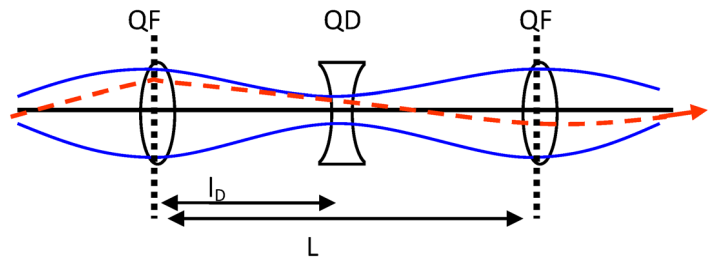

**Fig. 8:** Schematic drawing of a symmetric FODO cell

The task now is to calculate the Twiss parameters $\alpha, \beta, \gamma$ of such a cell. For convenience we will start our calculations in the middle of the focusing quadrupole. To start with the most simple configuration the drift spaces between the two quadrupole magnets will be really empty and of equal length.

In Fig. 9 the optical solution of typical arc structure is shown. Many FODO cells are connected with each other and the optical functions are calculated with a beam optics program. The plot shows the β function in both planes and below the position of the magnet lenses, the lattice. The solid line represents the horizontal $\beta$ function, the dashed one the vertical: it is evident, that the solution for both $\beta$ is periodic.

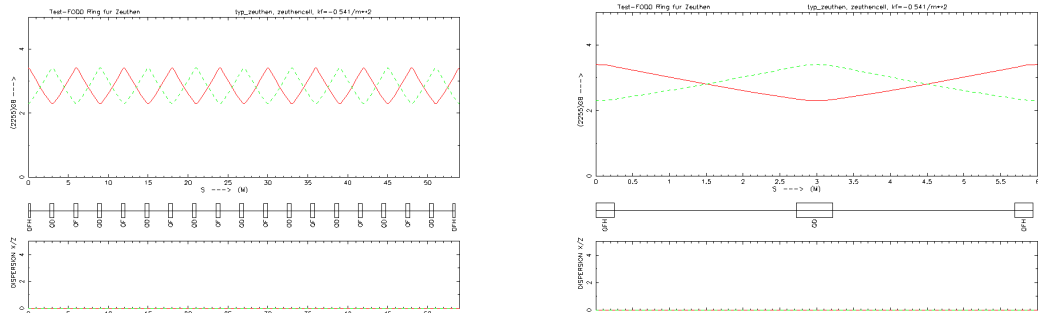

**Fig. 9**: Left: Beam optics in a periodic structure of nine FODO cells. Right: The corresponding situation calculated for a single cell.

The question now is can we understand what the optics code is doing? For this purpose we refer to one single cell.

Qualitatively spoken it is already clear from Fig. 9 that the horizontal $\beta_x$ reaches its maximum value at the centre of the (horizontal) focusing quadrupoles and its minimum value at the defocusing lenses. For the vertical function $\beta_y$ the statement holds vice versa. Table 2 below shows the numerical result of the optics code.

The $\alpha$ function in the middle of the quadrupole is indeed zero and, as $\alpha(s) = -\beta'(s)/2$, the $\beta$ function is maximum or minimum at that position.

The phase advance of this single cell is $\varphi=45^0$ and if the ring were built out of one such cell the corresponding working point working point would be $Q_x = \int \varphi \, ds \, /2\pi = 0.125$.

**Table 2:** Result of an optics calculation for a FODO lattice

| Element | $l$ (m) | $k$ (1/m$^2$) | $\beta_x$ (m) | $\alpha_x$ | $\varphi_x$ (rad) | $\beta_y$ (m) | $\alpha_y$ | $\varphi_y$ (rad) |
|---|---|---|---|---|---|---|---|---|
| Start | 0 | – | 11.611 | 0 | 0 | 5.295 | 0 | 0 |
| QFH | 0.25 | –0.0541 | 11.228 | 1.514 | 0.0110 | 5.488 | –0.78 | 0.0220 |
| QD | 3.251 | 0.0541 | 5.4883 | –0.78 | 0.2196 | 11.23 | 1.514 | 0.2073 |
| QFH | 6.002 | –0.0541 | 11.611 | 0 | 0.3927 | 5.295 | 0 | 0.3927 |
| End | 6.002 | – | 11.611 | 0 | 0.3927 | 5.295 | 0 | 0.3927 |

A final statement: As we have chosen equal quadrupole strengths in both planes, i.e. $k_x = -k_y$, and uniform drift spaces between the quadrupoles, the lattice is called a symmetric FODO cell and we obtain symmetric optical solutions in the two transverse planes.

In linear beam optics the transfer matrix of a number of optical elements is given by the product of the matrices of the single elements that have been introduced in Eq. (6). In our case we get

$$M_{FODO} = M_{QFH} \cdot M_{LD} \cdot M_{QD} \cdot M_{LD} \cdot M_{QFH}. \tag{10}$$

It has to be pointed out that as we have decided to start the calculation in the middle of a quadrupole magnet, the corresponding matrix has to take that into account: it will be that of a half quadrupole. Putting in the numbers for the length and strength, $k = \pm 0.54102/m^2$, $l_q = 0.5$ m, $l_d = 2.5$ m we get

$$M_{drift} = \begin{pmatrix} 0.707 & 8.206 \\ -0.061 & 0.707 \end{pmatrix}.$$

As we will see now, this matrix describes uniquely the optical properties of the lattice and defines the beam parameters.

### 4.1.1 The most important point: stability of the motion

The trace of $M$ gives

$$|trace(M_{FODO})| = 1.415 < 2.$$

A lattice built out of such FODO cells therefore would give stable conditions for the particle motion. I want to point out, however, that in the case where we introduce new parts in the lattice we will have to go through that calculation again, as we will see later.

In addition, the matrix can be used to determine the optical parameters of the system.

### 4.1.2 Phase advance per cell

Writing $M$ as a function of the Twiss parameters $\alpha, \beta, \gamma$ and the phase advance $\varphi$ we obtain

$$M(s) = \begin{pmatrix} \cos\varphi + \alpha_s \sin\varphi & \beta_s \sin\varphi \\ -\gamma_s \sin\varphi & \cos\varphi - \alpha_s \sin\varphi \end{pmatrix} \tag{11}$$

and we immediately see that

$$\cos\varphi = \frac{1}{2}|trace(M)| = 0.707$$

or $\varphi = 45°$, which corresponds to the working point of 0.125 calculated above.

### 4.1.3 The α and β functions

The α and β functions are calculated in a similar way. For $\beta$ we use the relation

$$\beta = \frac{M_{1,2}}{\sin\varphi} = 11.611 \text{ m}$$

and we get the parameter $\alpha$ through the expression

$$\alpha = \frac{M_{1,1} - \cos\varphi}{\sin\varphi} = 0.$$

To complete this first look at the optical properties of a lattice cell; I want to give a rule of thumb for the working point: defining an average $\beta$-function of the ring we put

$$\int \frac{ds}{\beta} \approx \frac{L}{\overline{\beta}}$$

with $L = 2\pi\overline{R}$ and $\overline{R}$ being the average bending radius of the ring (which is *not* the bending radius ρ of the dipole magnets) we can write for the working point $Q$

$$Q = N \cdot \frac{\varphi_c}{2\pi} = \frac{1}{2\pi} \oint \frac{ds}{\beta(s)} \approx \frac{1}{2\pi} \frac{2\pi\overline{R}}{\overline{\beta}}$$

where $N$ is the number of cells and $\varphi_C$ denotes the phase advance per cell. So we get

$$Q \approx \frac{\overline{R}}{\overline{\beta}}.$$

A rough estimate for the working point is obtained by the ratio of the mean radius of the ring and the average $\beta$-function of the lattice.

## 4.2 Thin lens approximation

As we have seen, a first estimate of the parameters of a lattice can and should be done at the beginning of the design of a magnet lattice.

If we want to have fast answers and require only rough estimates we can even do this a little bit easier: Under certain circumstances the matrix of a focusing element can be written in the so-called thin lens approximation.

Given, for example, the matrix of a focusing lens

$$M_{QF} = \begin{pmatrix} \cos(\sqrt{K} \cdot s) & \frac{1}{\sqrt{K}}\sin(\sqrt{K} \cdot s) \\ -\sqrt{K} \cdot \sin(\sqrt{K} \cdot s) & \cos(\sqrt{K} \cdot s) \end{pmatrix}$$

we can simplify the trigonometric terms if the focal length of the quadrupole magnet is much larger than the length of the lens:

$$\text{if} \quad f = \frac{1}{kl_q} \gg l_q,$$

the transfer matrix can be approximated using $kl_Q$ = const, $l_Q \to 0$ and we get,

$$M_{QF} = \begin{pmatrix} 1 & 0 \\ -1/f & 1 \end{pmatrix}.$$

Applying this concept to Eq. (10) we can rewrite the expression of a FODO cell in the thin lens approximation. To reduce the effort we start again in the middle of the focusing quadrupole and calculate the product matrix up to the middle of the defocusing quadrupole to obtain the matrix for half the cell:

$$M_{\text{half cell}} = M_{QDH} \cdot M_{LD} \cdot M_{QFH},$$

$$M_{\text{half cell}} = \begin{pmatrix} 1 & 0 \\ 1/\tilde{f} & 1 \end{pmatrix} \cdot \begin{pmatrix} 1 & l_d \\ 0 & 1 \end{pmatrix} \cdot \begin{pmatrix} 1 & 0 \\ -1/\tilde{f} & 1 \end{pmatrix},$$

$$M_{\text{half cell}} = \begin{pmatrix} 1 - l_d/\tilde{f} & l_d \\ -l_d/\tilde{f}^2 & 1 + l_d/\tilde{f} \end{pmatrix}. \tag{12}$$

Note that the thin lens approximation implies that $l_Q \to 0$, therefore the drift between the magnets has to be $l_D = L/2$. As we are dealing now with half quadrupoles we have set for the focal length of a half quadrupole $\tilde{f} = 2f$.

The second half of the cell we get by replacing $\tilde{f}$ by $-\tilde{f}$ and the matrix for the complete FODO in thin lens approximation is

$$M_{\text{half cell}} = \begin{pmatrix} 1 + l_d/\tilde{f} & l_d \\ -l_d/\tilde{f}^2 & 1 - l_d/\tilde{f} \end{pmatrix} \cdot \begin{pmatrix} 1 - l_d/\tilde{f} & l_d \\ -l_d/\tilde{f}^2 & 1 + l_d/\tilde{f} \end{pmatrix}$$

or multiplying out

$$M_{\text{half cell}} = \begin{pmatrix} 1 - \frac{2l_d^2}{\tilde{f}^2} & 2l_d\left(1 + \frac{l_d}{\tilde{f}}\right) \\ 2\left(\frac{l_d^2}{\tilde{f}^3} - \frac{l_d}{\tilde{f}^2}\right) & 1 - \frac{2l_d^2}{\tilde{f}^2} \end{pmatrix}. \tag{13}$$

The matrix is now much easier to handle than the equivalent formulae (4) and (8) and the approximation is in general not bad.

Going briefly again through the calculation of the optics parameters we get immediately according to (11) and (13)

$$\cos\varphi = 1 - \frac{2l_d^2}{\tilde{f}^2}$$

and with a little bit of trigonometric gymnastics

$$1 - 2\sin^2\frac{\varphi}{2} = 1 - \frac{2l_d^2}{\tilde{f}^2}$$

we can simplify this expression and get

$$\sin\frac{\varphi}{2} = \frac{l_d}{\tilde{f}} = \frac{L_{\text{cell}}}{2\tilde{f}}$$

and finally

$$\sin\frac{\varphi}{2} = \frac{L_{\text{cell}}}{4f}. \qquad (14)$$

In thin lens approximation the phase advance of a FODO cell is given by the length of the cell $L_{\text{cell}}$ and the focal length of the quadrupole magnets $f$.

For the parameters of the example that was given above we get a phase advance per cell of $\varphi \approx 47.8°$ and in full analogy to the calculation presented before we calculate $\beta \approx 11.4$ m, which is very close to the result of the exact calculations ($\varphi = 45°$, $\beta = 11.6$ m).

### 4.2.1 *Stability of the motion*

In thin lens approximation the condition for stability $|trace(M)| < 2$ requires that

$$2 - \frac{4l_d^2}{\tilde{f}^2} < 2$$

or

$$f > \frac{L_{\text{cell}}}{4}.$$

We have got the important and simple result that for stable motion the focal length of the quadrupole lenses in the FODO has to be larger than a quarter of the length of the cell.

### 4.3 Scaling optical parameters of a lattice cell

After the discussion on stability in a lattice cell and after first estimates and calculations of the optical functions $\alpha$, $\beta$, $\gamma$, and $\varphi$, we would like to concentrate now a little bit more on a detailed analysis of a FODO concerning these parameters.

As we have seen we can calculate the $\beta$ function that corresponds to the periodic solution, provided that we know the strength and length of the focusing elements in the cell. But can we optimize somehow?

In other words, for a given lattice, what would be the ideal magnet strength to get the smallest beam dimensions? To answer this question we will go back to the transfer matrix of half a FODO cell as indicated in Eq. (12), i.e. the transfer from the middle of a QF quadrupole to the middle of a QD (see Fig. 8).

From linear beam optics we know that the transfer matrix between two points in a lattice can be expressed not only as a function of the focusing properties of the elements in that section of the ring, but in an equivalent way as a function of the optical parameters between the two reference points. For a full turn or, within a periodic lattice for one period, we have used that relation already in Eq. (7).

The general expression, in the non-periodic case, reads [4]

$$M_{1\to 2} = \begin{pmatrix} \dfrac{\sqrt{\beta_2}}{\sqrt{\beta_1}}\cos\Delta\varphi + \alpha_1 \sin\Delta\varphi & \sqrt{\beta_1\beta_2}\sin\Delta\varphi \\ \dfrac{(\alpha_1-\alpha_2)\cos\Delta\varphi - (1+\alpha_1\alpha_2)\sin\Delta\varphi}{\sqrt{\beta_1\beta_2}} & \dfrac{\sqrt{\beta_1}}{\sqrt{\beta_2}}\cos\Delta\varphi - \alpha_2 \sin\Delta\varphi \end{pmatrix}. \tag{15}$$

The indices refer to the starting point $s_1$ and the end point $s_2$ in the ring and $\Delta\varphi$ is the phase advance between these points. It is evident that this matrix is reduced to the form given in Eq. (7) if the periodic conditions $\beta_1 = \beta_2$, $\alpha_1 = \alpha_2$ are fulfilled.

In the middle of the focusing quadrupole we know already that $\beta$ reaches its highest value and in the middle of the defocusing magnet its lowest one and so the $\alpha$ functions at those positions are zero. Therefore, the transfer matrix that takes us from the centre of a QF to the centre of a QD can be written in the form

$$M = \begin{pmatrix} C & S \\ C' & S' \end{pmatrix} = \begin{pmatrix} \dfrac{\sqrt{\breve{\beta}}}{\sqrt{\hat{\beta}}}\cos\Delta\varphi & \sqrt{\hat{\beta}\breve{\beta}}\sin\Delta\varphi \\ \dfrac{-1}{\sqrt{\hat{\beta}\breve{\beta}}}\sin\Delta\varphi & \dfrac{\sqrt{\hat{\beta}}}{\sqrt{\breve{\beta}}}\cos\Delta\varphi \end{pmatrix}.$$

Using this expression and putting for the matrix elements the terms that we have developed in thin lens approximation in Eq. (12) we get

$$\frac{\sqrt{\hat{\beta}}}{\sqrt{\breve{\beta}}} = \frac{S'}{C} = \frac{1 + l_d/\tilde{f}}{1 - l_d/\tilde{f}} = \frac{1+\sin\dfrac{\varphi}{2}}{1-\sin\dfrac{\varphi}{2}},$$

$$\hat{\beta}\cdot\breve{\beta} = \frac{-S}{C'} = \tilde{f}^2 = \frac{l_d^2}{\sin^2\dfrac{\varphi}{2}},$$

where we have set $\Delta\varphi = \varphi/2$ for the phase advance of half the FODO cell. The two expressions for $\hat{\beta}$ and $\breve{\beta}$ can be combined to calculate both parameters

$$\hat{\beta} = \frac{\left(1+\sin\dfrac{\varphi}{2}\right)L}{\sin\varphi}, \qquad \breve{\beta} = \frac{\left(1-\sin\dfrac{\varphi}{2}\right)L}{\sin\varphi}. \tag{16}$$

We get the simple result that the maximum (and minimum) value of the $\beta$ function and therefore the maximum dimension of the beam in the cell are determined by the length $L$ and the phase advance $\varphi$ of the complete cell.

Figure 10 shows qualitatively a three-dimensional picture of a proton bunch for typical conditions in a storage ring. The bunch length is about 30 cm and determined by the momentum spread and the RF potential [6].

The values of $\hat{\beta}$ and $\check{\beta}$, as determined by the cell characteristics, are typically 80 and 40 m, respectively.

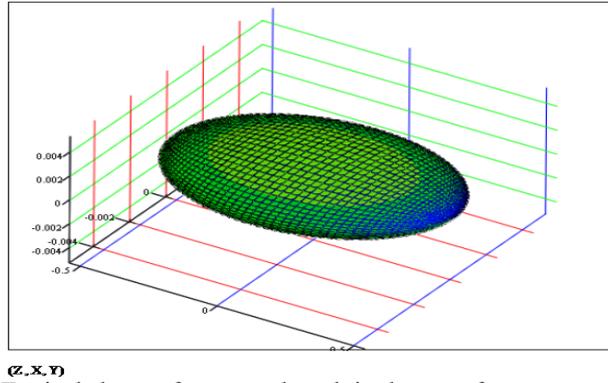

**Fig. 10:** Typical shape of a proton bunch in the arc of a proton storage ring

### 4.3.1 Optimization of the FODO phase advance

From Eq. (16) we see that, given the length of the FODO, the maximum $\beta$ depends only on the phase advance per cell. Therefore, we may ask the question whether there is an optimum phase that leads to the smallest beam dimension.

If we assume a Gaussian particle distribution in the transverse plane and denote the beam emittance by $\varepsilon$, the transverse beam dimension $\sigma$ is given by

$$\sigma = \sqrt{\varepsilon \beta}.$$

In a typical high-energy proton ring $\varepsilon$ is of the order of $10^{-9}$ m · rad (e.g. the LHC proton ring at $E = 7000$ GeV, $\varepsilon \approx 0.5 \times 10^{-9}$ m · rad) and as the typical $\beta$ functions are about $\beta \approx 80$–$120$ m in the arc the resulting beam dimension is roughly a millimetre. At the interaction point of two counter-rotating beams even beam dimensions of the order of micrometre are obtained.

Figure 11 shows the result of a beam scan which is used to measure the transverse beam dimension.

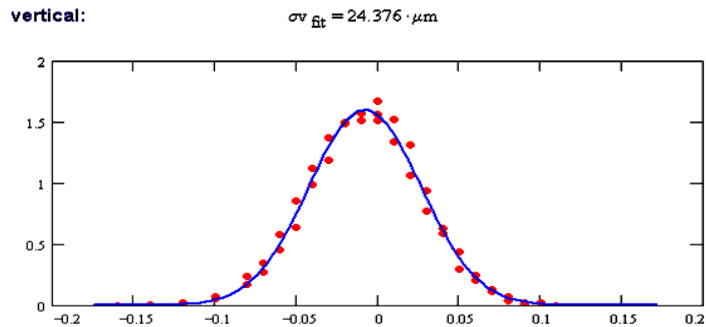

**Fig. 11:** Transverse beam profile of a proton bunch at the interaction point. The measurement is performed by scanning the colliding beams against each other.

In general for a proton beam both emittances are equal, $\varepsilon_x \approx \varepsilon_y$. In that sense a proton beam is 'round' even if the varying $\beta$ function along the lattice leads to beam dimensions in the two transverse planes that can be quite different.

To optimize the beam size in the case of a proton ring therefore means the search for a minimum of the beam radius:

$$r^2 = \varepsilon_x \beta_x + \varepsilon_y \beta_y$$

and therefore to optimize the sum of the maximum and minimum $\beta$ functions at the same time:

$$\hat{\beta} + \breve{\beta} = \frac{\left(1 + \sin\frac{\varphi}{2}\right)L}{\sin\varphi} + \frac{\left(1 - \sin\frac{\varphi}{2}\right)L}{\sin\varphi}. \tag{17}$$

The optimum phase $\varphi$ is obtained by the derivative

$$\frac{d}{d\varphi}(\hat{\beta} + \breve{\beta}) = \frac{d}{d\varphi}\left(\frac{2L}{\sin\varphi}\right) = 0,$$

which gives

$$\frac{L}{\sin^2\varphi}\cos\varphi = 0 \quad \rightarrow \quad \varphi = 90°.$$

Concerning the aperture requirement of the cell, a phase advance of $\varphi = 90°$ is the best value in a proton ring. The plot of Fig. 12 shows the sum of the two $\beta$ (Eq. (17)) as a function of the phase $\varphi$ in the range of $\varphi = 0$–$180°$.

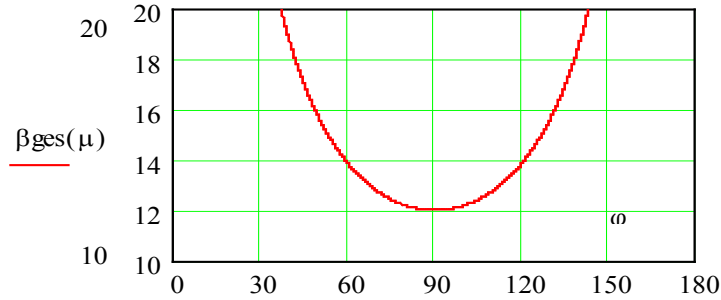

**Fig. 12:** Sum of the horizontal and vertical β function as a function on the phase advance

It has to be pointed out that the optimization of the beam radii can be a critical issue in the accelerator design: large beam dimensions need big apertures of the quadrupole and dipole magnets in the ring. Running the machine at the highest energy therefore can lead to limitations in the focusing power as the gradient of a quadrupole lens scales as the inverse of its squared aperture radius $r_0^2$ and increases the cost for the magnet lenses. Therefore, it is recommended not to tune the lattice too far away from the ideal phase advance.

Here for completeness, I have to make a short remark on electron machines: unlike to the situation in proton rings, electron beams are flat in general. Owing to the mechanism of the synchrotron radiation [7] the vertical emittance of an electron or positron beam is only a small fraction of the horizontal one $\varepsilon_y \approx 1$–$10\%\ \varepsilon_x$. For the optimization of the phase advance the calculation can and should be restricted to the horizontal plane only and the condition for smallest beam dimension is

$$\frac{d}{d\varphi}(\hat{\beta}) = \frac{d}{d\varphi}\frac{\left(1 + \sin\frac{\varphi}{2}\right)L}{\sin\varphi} = 0 \quad \rightarrow \quad \varphi = 76°.$$

Figure 13 shows the horizontal and vertical $\beta$ as a function of $\varphi$ in that case. In an electron ring the typical phase advance is in the range of $\varphi \approx 30$–$90°$.

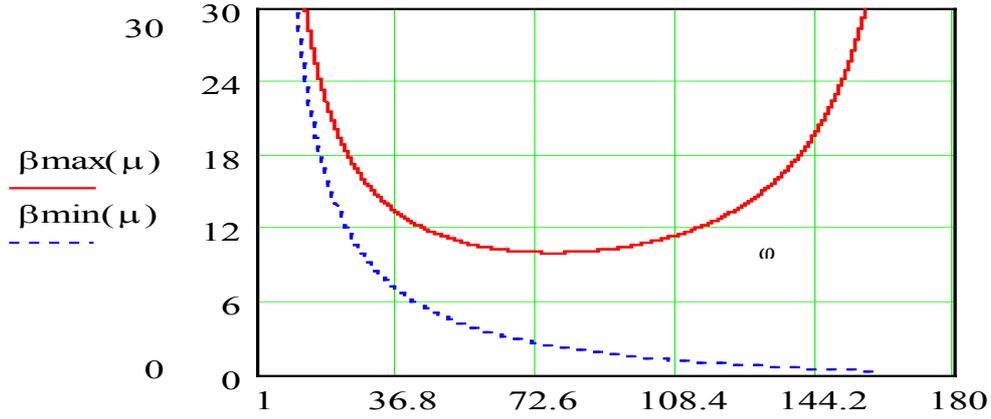

**Fig. 13:** Horizontal and vertical $\beta$ in a FODO cell as a function of the phase advance $\varphi$

### 4.4 Dispersion in a FODO lattice

In the design and in the description of the optical parameters of a magnet lattice we have restricted ourselves until now to the case where all particles have ideal momentum. In doing so, we have followed the usual path of typical books about linear beam dynamics. But, in general, the energy (or momentum) of the particles stored in a ring is not ideal but rather distributed around an average value and the individual particle momentum $p$ will deviate from the ideal momentum $p_0$ of the beam.

We know from linear beam optics that the differential equation for the transverse movement gets an additional term if the momentum deviation is not zero, $\Delta p/p \neq 0$. We get an inhomogeneous equation of motion

$$x'' + K(s) \cdot x = \frac{1}{\rho} \frac{\Delta p}{p_0}. \tag{18}$$

The left-hand side of (18) is the same as in the homogeneous case Eq. (5) and the parameter $K$ describes as usual the focusing strength of the lattice element at position $s$ in the ring. The general solution of Eq. (18) is the sum of the complete solution $x_h$ of the homogeneous equation and a special solution of the inhomogeneous one $x_i$:

$$x_h'' + K(s) \cdot x_h = 0,$$

$$x_i'' + K(s) \cdot x_i = \frac{1}{\rho} \frac{\Delta p}{p_0}.$$

For convenience the special solution $x_i$ can be normalized to the momentum error $\Delta p/p$ and one obtains the so-called dispersion function $D(s)$

$$x_i(s) = D(s) \frac{\Delta p}{p}. \tag{19}$$

Starting as before from the initial conditions $x_0$ and $x_0'$, the general solution for the particle trajectory now reads

$$x(s) = C(s)x_0 + S(s)x_0' + D(s)\frac{\Delta p}{p}$$

or including the expression for the angle $x'(s)$

$$\begin{pmatrix} x \\ x' \end{pmatrix}_s = \begin{pmatrix} C & S \\ C' & S' \end{pmatrix} \cdot \begin{pmatrix} x \\ x' \end{pmatrix}_0 + \frac{\Delta p}{p} \begin{pmatrix} D \\ D' \end{pmatrix}.$$

In general one extends the matrix to include the second term and can write

$$\begin{pmatrix} x \\ x' \\ \Delta p/p \end{pmatrix}_s = \begin{pmatrix} C & S & D \\ C' & S' & D' \\ 0 & 0 & 1 \end{pmatrix} \cdot \begin{pmatrix} x \\ x' \\ \Delta p/p \end{pmatrix}_0 .$$

The dispersion function $D(s)$ is (obviously) defined by the focusing properties of the lattice and the bending strength of the dipole magnets $1/\rho$ and it can be shown that [5]

$$D(s) = S(s) * \int \frac{1}{\rho(\tilde{s})} C(\tilde{s}) \, d\tilde{s} - C(s) * \int \frac{1}{\rho(\tilde{s})} S(\tilde{s}) \, d\tilde{s}. \tag{20}$$

The variable $s$ refers to the position where the dispersion is obtained (or measured if you like) and the integration has to be performed over all places $\tilde{s}$ where a non-vanishing term $1/\rho$ exists (in general, in the dipole magnets of the ring).

As an example, the 2 × 2 matrix for a drift space is given by

$$M_{\text{drift}} = \begin{pmatrix} C & S \\ C' & S' \end{pmatrix} = \begin{pmatrix} 1 & l \\ 0 & 1 \end{pmatrix}.$$

As there are no dipoles in the drift the $1/\rho$ term in Eq. (20) is zero and we obtain the extended 3 × 3 matrix

$$M_{\text{drift}} = \begin{pmatrix} 1 & l_d & 0 \\ 0 & 1 & 0 \\ 0 & 0 & 1 \end{pmatrix}.$$

To calculate the dispersion in a FODO cell we refer again to the thin lens approximation that had already been used for the calculation of the $\beta$ functions. The matrix for a half cell has been derived above (see Eq. (12)). Again we want to point out that in thin lens approximation the length $l$ of the drift is just half the length of the cell, as the quadrupole lenses have zero length:

$$M_{\text{half cell}} = \begin{pmatrix} 1 - l_d/\tilde{f} & l_d \\ -l_d/\tilde{f}^2 & 1 + l_d/\tilde{f} \end{pmatrix}.$$

Using this expression we can calculate the terms $D$, $D'$ of the 3 × 3 matrix

$$D(s) = S(s) * \int \frac{1}{\rho(\tilde{s})} C(\tilde{s}) \, d\tilde{s} - C(s) * \int \frac{1}{\rho(\tilde{s})} S(\tilde{s}) \, d\tilde{s},$$

where the integral is taken over the length $l_d$ of the half cell

$$D(l_d) = l_d \frac{1}{\rho} \int_0^{l_d} \left(1 - \frac{s}{\tilde{f}}\right) ds - \left(1 - \frac{l_d}{\tilde{f}}\right) \frac{1}{\rho} \int_0^{l_d} s \, ds,$$

$$D(l_d) = \frac{l_d}{\rho}\cdot\left(1-\frac{l_d^2}{2\tilde{f}}\right) - \left(1-\frac{l_d}{\tilde{f}}\right)\cdot\frac{1}{\rho}\frac{l_d^2}{2} = \frac{l_d^2}{\rho} - \frac{l_d^3}{2\tilde{f}\rho} - \frac{l_d^2}{2\rho} + \frac{l_d^3}{2\tilde{f}\rho},$$

$$D(l_d) = \frac{l_d^2}{2\rho}.$$

In an analogous way one derives the expression for $D'$

$$D'(l_d) = \frac{l_d}{\rho}\left(1+\frac{l_d}{2\tilde{f}}\right)$$

and we get the complete matrix for a FODO half cell

$$M_{\text{half cell}} = \begin{pmatrix} C & S & D \\ C' & S' & D' \\ 0 & 0 & 1 \end{pmatrix} = \begin{pmatrix} 1-{l_d}/{\tilde{f}} & l_d & \dfrac{l_d^2}{2\rho} \\ -{l_d}/{\tilde{f}^2} & 1+{l_d}/{\tilde{f}} & \dfrac{l_d}{\rho}\left(1+\dfrac{l_d}{2\tilde{f}}\right) \\ 0 & 0 & 1 \end{pmatrix}.$$

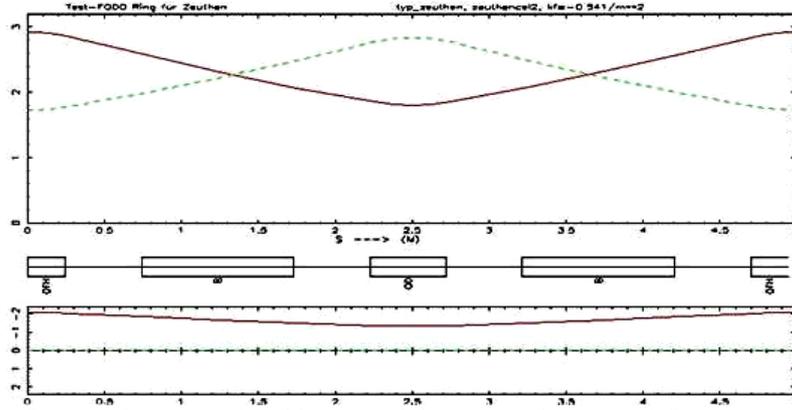

**Fig. 14:** Beta function (upper part) and horizontal dispersion (lower part) function in a FODO cell

Now we know that due to symmetry the dispersion in a FODO lattice reaches its maximum value in the centre of a QF quadrupole and its minimum in a QD, as for example is shown in Fig. 14 where in addition to the β function the dispersion is shown in the lower part of the figure.

Therefore, we get the boundary conditions for the transformation from a QF to a QD lens

$$\begin{pmatrix} \check{D} \\ 0 \\ 1 \end{pmatrix} = M_{\text{half cell}} \cdot \begin{pmatrix} \hat{D} \\ 0 \\ 1 \end{pmatrix},$$

which can be used to calculate the dispersion at these locations:

$$\check{D} = \hat{D}\left(1-\frac{l_d}{\tilde{f}}\right) + \frac{l_d^2}{2\rho},$$

$$0 = -\frac{l_d}{\tilde{f}^2} \cdot \hat{D} + \frac{l_d}{\rho} \cdot \left(1 + \frac{l_d}{2\tilde{f}}\right).$$

Remember that we have to use the focal length of a half quadrupole
$$\tilde{f} = 2f$$

and that the phase advance is given by
$$\sin\frac{\varphi}{2} = \frac{L_{\text{cell}}}{2\tilde{f}}.$$

For the maximum dispersion in the middle of a focusing quadrupole and for the minimum dispersion in the middle of a defocusing lens we obtain the expressions

$$\hat{D} = \frac{l_d^2}{\rho} \cdot \frac{1 + \frac{1}{2}\sin\frac{\varphi}{2}}{\sin^2\varphi}, \qquad \check{D} = \frac{l_d^2}{\rho} \cdot \frac{1 - \frac{1}{2}\sin\frac{\varphi}{2}}{\sin^2\varphi}. \tag{21}$$

It is interesting to note that the dispersion depends only on the half length $l$ of the cell, the bending strength of the dipole magnet $1/\rho$, and the phase advance $\varphi$. The dependence of $D$ on the phase advance is shown in the plot of Fig. 15. Both values, $D_{\max}$ and $D_{\min}$, are decreasing for an increasing phase $\varphi$ (which is just another way of saying 'for increasing focusing strength', as $\varphi$ depends on the focusing strength of the quadrupole magnets).

To summarize these considerations I would like to make the following remarks:

i) small dispersion needs strong focusing and therefore large phase advance;

ii) there is however an optimum phase advance concerning the best (i.e. smallest) value of the $\beta$ function;

iii) even more, the stability criterion limits the choice of the phase advance per cell.

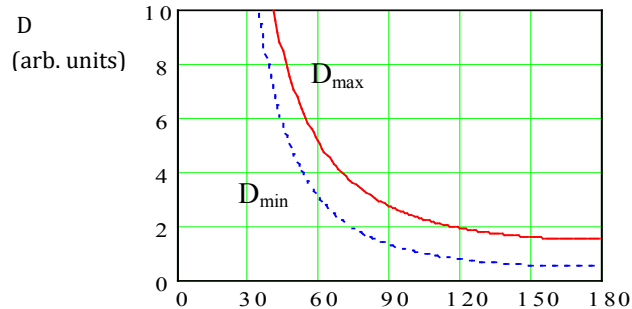

**Fig. 15:** Dispersion at the focusing and defocusing quadrupole lens in a FODO as a function of the phase $\varphi$

In general, therefore, one has to find a compromise for the focusing strength in a lattice that takes into account the stability of the motion, the $\beta$ function in both transverse planes and the dispersion function. In a typical high-energy machine this optimization is not too difficult as the dispersion does not have too much of an impact on the beam parameters (as long as it is compensated for at the interaction point of the two beams).

In synchrotron light sources, however, the beam emittance is usually the parameter that has to be optimized (that means in nearly all cases *minimized*) and as the emittance depends on the dispersion $D$ in an electron storage ring the dispersion function and its optimization is of greatest importance in these machines.

In an electron ring the horizontal beam emittance is given by the expression

$$\varepsilon_x = \frac{55}{32\sqrt{3}} \frac{\hbar}{mc} \gamma^2 \frac{\left\langle \frac{1}{R^3} H(s) \right\rangle}{J_x \left\langle \frac{1}{R^2} \right\rangle},$$

where the function $H(s)$ is defined by

$$H(s) = \gamma D^2 + 2\alpha DD' + \beta D'^2.$$

The optimization of *H(s)* in a magnet structure for electron beams is a subject of its own and an introduction into that field of the so-called low-emittance lattices can be found, for example, in [8].

### 4.5 Orbit distortions in a periodic lattice

The lattice that we have designed so far consists only of a small number of basic elements: bending magnets that define the geometry of the circular accelerator and for a given particle momentum the size of the machine, quadrupole lenses defining the phase advance of the single particle trajectory and through this parameter the beam dimensions and the stability of the motion.

Now it is time to fill the empty spaces in the lattice cell with some other useful components, which means we have to talk about the '*Os*' of the FODO.

Nobody is perfect and this statement also holds for storage rings. In the linear beam optics lecture, field errors in storage rings have been discussed and we know that in the case of a dipole magnet an error of the bending field is described by an additional kick $\delta$ (typically measured in mrad) on the particles

$$\delta = \frac{ds}{\rho} = \frac{\int B \, ds}{p/e}.$$

The beam is oscillating in the corresponding plane and the resulting amplitude of the orbit is

$$x(s) = \frac{\sqrt{\beta(s)}}{2 \sin \pi Q} \oint \sqrt{\beta(\tilde{s})} \frac{1}{\rho(\tilde{s})} \cos(|\varphi(\tilde{s}) - (s)| - \pi Q) \, d\tilde{s}. \quad (22)$$

It is given by the $\beta$ function at the place of the dipole magnet $\beta(\tilde{s})$ and its bending strength $1/\rho$, and the $\beta$ function at the observation point in the lattice $\beta(s)$. For the lattice designer, that means that if a correction magnet has to be installed in the lattice cell, it should be placed at a location where $\beta$ is high in the corresponding plane.

At the same time Eq. (22) tells us that the amplitude of an orbit distortion is highest at the place where $\beta$ is high and this is the place where beam position monitors have to be located to measure the orbit distortion precisely. In practice, therefore, both beam position monitors and orbit correction coils are located at places in the lattice cell where the $\beta$ function in the considered plane is large, i.e. close to the corresponding quadrupole lens.

### 4.6 Chromaticity in a FODO cell

The chromaticity $Q'$ describes an optical error of a quadrupole lens in an accelerator: for a given magnetic field, i.e. the gradient of the quadrupole magnet, particles with smaller momentum will feel a stronger focusing force and vice versa.

The chromaticity $Q'$ relates the resulting tune shift to the relative momentum error of the particle

$$\Delta Q = Q' \cdot \frac{\Delta p}{p}$$

and as it is a consequence of the focusing properties of the quadrupole magnets it is given by the characteristics of the lattice. For small momentum errors $\Delta p/p$ the focusing parameter $k$ can be written as

$$k(p) = \frac{g}{p/e} = g \cdot \frac{e}{p_0 + \Delta p},$$

where $g$ denotes the gradient of the quadrupole lens, $p_0$ the design momentum and the term $\Delta p$ refers to the momentum error. If $\Delta p$ is small as we have assumed, we can write

$$k(p) = g \cdot \frac{e}{p_0}\left(1 - \frac{\Delta p}{p}\right) = k + \Delta k.$$

This describes a quadrupole error

$$\Delta k = -k_0 \cdot \frac{\Delta p}{p}$$

and leads to a tune shift of

$$\Delta Q = \frac{1}{4\pi}\int \Delta k \cdot \beta(s)\,\mathrm{d}s,$$

$$\Delta Q = \frac{-1}{4\pi}\frac{\Delta p}{p}\int k_0 \cdot \beta(s)\,\mathrm{d}s.$$

By definition, the chromaticity $Q'$ of a lattice is therefore given by

$$\Delta Q = \frac{-1}{4\pi}\int \beta(s)k(s)\,\mathrm{d}s. \qquad (23)$$

Let us assume now that the accelerator consists of $N$ identical FODO cells. Then, replacing the $\beta(s)$ by its maximum value at the focusing, and its minimum value at the defocusing quadrupoles we can approximate the integral in Eq. (23) by the sum

$$Q' = \frac{-1}{4\pi} N \cdot \frac{\hat{\beta} - \check{\beta}}{f_q},$$

where $f_q = 1/(k*l)$ denotes as usual the focal length of the quadrupole magnet and

$$Q' = \frac{-1}{4\pi} N \frac{1}{f_q}\left\{\frac{L\left(1+\sin\frac{\varphi}{2}\right) - L\left(1-\sin\frac{\varphi}{2}\right)}{\sin\varphi}\right\}, \qquad (24)$$

where we have used the expressions from Eq. (16) for $\hat{\beta}$ and $\check{\beta}$. With some useful trigonometric transformations such as

$$\sin x = 2\sin\frac{x}{2}\cdot\cos\frac{x}{2},$$

we can transform the right-hand side of (24) to obtain

$$Q' = \frac{-1}{4\pi} N \frac{1}{f_q} \frac{L\left(\sin\frac{\varphi}{2}\right)}{\sin\frac{\varphi}{2} \cdot \cos\frac{\varphi}{2}}$$

or for one single cell, $N = 1$,

$$Q' = \frac{-1}{4\pi} \frac{1}{f_q} \frac{L\left(\sin\frac{\varphi}{2}\right)}{\sin\frac{\varphi}{2} \cdot \cos\frac{\varphi}{2}}.$$

Remembering the relation

$$\sin\frac{\varphi}{2} = \frac{L}{4f_q},$$

we obtain a surprisingly simple result for the chromaticity contribution of a single FODO cell:

$$Q'_{\text{cell}} = \frac{-1}{\pi} \tan\frac{\varphi}{2}.$$

## 5  Lattice insertions

We have seen in Fig. 8 that the lattice of a typical machine for the acceleration of high-energy particles consists of two quite different parts: the arcs that are built out of a number of identical cells and the straight sections that connect them and that house complicated systems such as dispersion suppressors, mini beta insertions, or high-energy particle detectors.

### 5.1  Drift space

To get a first insight into the design of lattice insertions I would like to start with some comments concerning a simple drift space embedded in a normal lattice structure.

What will happen to the beam parameters $\alpha$, $\beta$, and $\gamma$ if we stop focusing for a while? The transfer matrix for the Twiss parameters from a point '0' to position '$s$' in a lattice is given by the formula

$$\begin{pmatrix} \beta \\ \alpha \\ \gamma \end{pmatrix}_s = \begin{pmatrix} C^2 & -2SC & S^2 \\ -CC' & SC' + CS' & -SS' \\ C'^2 & -2S'C' & S'^2 \end{pmatrix} \cdot \begin{pmatrix} \beta \\ \alpha \\ \gamma \end{pmatrix}_0,$$

where the cosine and sine functions $C$ and $S$ are given by the focusing properties of the lattice elements in between the two points.

For a drift space of length $s$ this is according to Eq. (6) as simple as

$$M_{\text{drift}} = \begin{pmatrix} C & S \\ C' & S' \end{pmatrix} = \begin{pmatrix} 1 & l \\ 0 & 1 \end{pmatrix}$$

and the optical parameters will develop as a function of $s$ in the following way

$$\beta(s) = \beta_0 - 2\alpha_0 s + \gamma_0 s^2,$$
$$\alpha(s) = \alpha_0 - \gamma_0 s, \quad (25)$$
$$\gamma(s) = \gamma_0.$$

We will have now a closer look at these relations.

### 5.1.1 *Location of a beam waist*

From the first equation we see immediately that if the drift space is long enough, even a convergent beam at position '0' will become divergent, as the term $\gamma_0 s^2$ is always positive. This is shown schematically in Fig. 16.

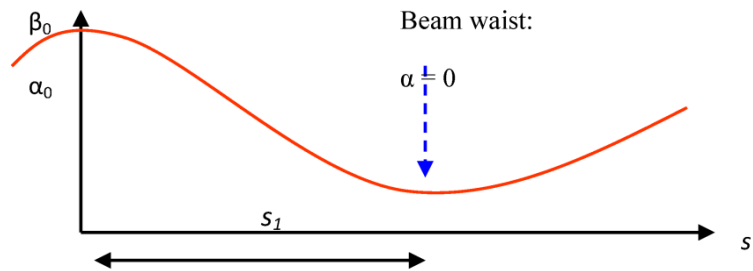

**Fig. 16:** Schematic drawing of a beam waist

Therefore, there will be a point $s_1$ in the drift, where the beam dimension is smallest, in other words where the beam envelope has a waist. The position of this waist can be calculated by requiring
$$\alpha(s_1) = 0$$
and the second equation of (25) then gives
$$\alpha_0 = \gamma_0 s_1$$
or
$$s_1 = \frac{\alpha_0}{\gamma_0}.$$

The position of the waist is given by the ratio of the α and γ functions at the beginning of the drift.

As the $\gamma$ parameter is constant in a drift and $\alpha$ is zero at the waist we can directly calculate the beam size that we get at the waist:
$$\gamma(s_1) = \gamma_0, \quad \alpha(s_1) = 0,$$
$$\beta(s_1) = \frac{1 + \alpha^2(s_1)}{\gamma(s_1)} = \frac{1}{\gamma_0}. \quad (26)$$

The $\beta$ function at the location of the waist is just given by the inverse of the gamma function at the beginning of the drift; a nice and simple scaling law.

### 5.1.2 *Beta function in a drift space*

It is worth thinking a little bit more about the behaviour of the Twiss parameters in a drift. Namely the scaling of $\beta$ with the length of the drift has a large impact on the design of high-energy machines. Let us assume that we are in the centre of a drift and that the situation is symmetric, that means the index

'0' refers to the position at the starting point, but now we want to have a left–right symmetric optics with respect to it, $\alpha_0 = 0$. From (25) we get at the starting point

$$\beta(s) = \beta_0 - 2\alpha_0 s + \gamma_0 s^2$$

and knowing already from (26) that at the waist $s = 0$ we have

$$\gamma_0 = \frac{1 + \alpha_0^2}{\beta_0} = \frac{1}{\beta_0},$$

we get $\beta$ as a function of the distance $s$ from the starting point:

$$\beta(s) = \beta_0 + \frac{s^2}{\beta_0}. \tag{27}$$

I would like to point out two facts in that context.

i) The relation (27) is a direct consequence of Liouville's theorem: the particle density in phase space is constant in an accelerator. In other words, if there are only conservative forces, the beam emittance $\varepsilon$ is constant, which leads immediately to the relation (27). As the conservation of $\varepsilon$ is a fundamental law there is no trick to avoid it and no way to overcome the increase of the beam dimension in a drift.

ii) The behaviour of $\beta$ in a drift has a strong impact on the design of a storage ring. As large beam dimensions have to be avoided, it means that large drift spaces are forbidden or at least very inconvenient. We will see in the next section that this is one of the major limitations of the luminosity of colliding beams in an accelerator.

At the beam waist we can derive another short relation that is often used for scaling of beam parameters: the beam envelope σ is given by the $\beta$ function and the emittance of the beam,

$$\sigma(s) = \sqrt{\varepsilon \cdot \beta(s)}$$

and the divergence σ' by

$$\sigma'(s) = \sqrt{\varepsilon \cdot \gamma(s)}.$$

Now, as $\gamma = (1+\alpha^2)/\beta$, wherever $\alpha = 0$ the beam envelope has a local minimum (i.e. a waist) or maximum. At that position the $\beta$ function is just the ratio of the beam envelope and the beam divergence at a waist

$$\beta(s) = \frac{\sigma(s)}{\sigma'(s)}.$$

If we cannot fight against Liouville's theorem, we can at least try to optimize its consequences. Equation (27) for $\beta$ in a symmetric drift can be used to find the starting value that gives the smallest beam dimension at the end of the drift of length $\ell$.

Setting

$$\frac{d\hat{\beta}}{d\beta_0} = 1 - \frac{l^2}{\beta_0^2} = 0$$

gives us the value of $\beta_0$ that leads to the smallest $\beta$ after a drift of length $l$:

$$\beta_0 = l. \tag{28}$$

For a starting value of $\beta_0 = l$, the maximum beam dimension at the end of the drift will be smallest and its value is just double the length of the drift.

$$\hat{\beta} = 2\beta_0 = 2l.$$

## 5.2 Mini beta insertions

The discussion of the last section has shown that the β function in a drift space can be chosen with respect to the length *l* to minimize the beam dimension and according to that the aperture requirements for vacuum chambers and magnets. In general, the value of β is of the order of some metres and the typical length of drift spaces in a lattice is of the same order.

However, the straight sections of a storage ring are often designed for the collision of two counter-rotating beams. The β functions at the collision points therefore are very small compared with their values in the arc cells. Typical values are more in the range of centimetres than of metres. Still the same scaling law (28) holds and the optimum length of such a drift would be for example $l \approx 55$ cm for the interaction regions of the two beams in the LHC.

Modern high-energy detectors in contrast are impressive devices that consist of many large components and they do not fit in a drift space of some centimetres. Figure 17 shows the ATLAS detector at the LHC. It is evident that for the installation of such a huge detector a special treatment of the storage ring lattice is needed.

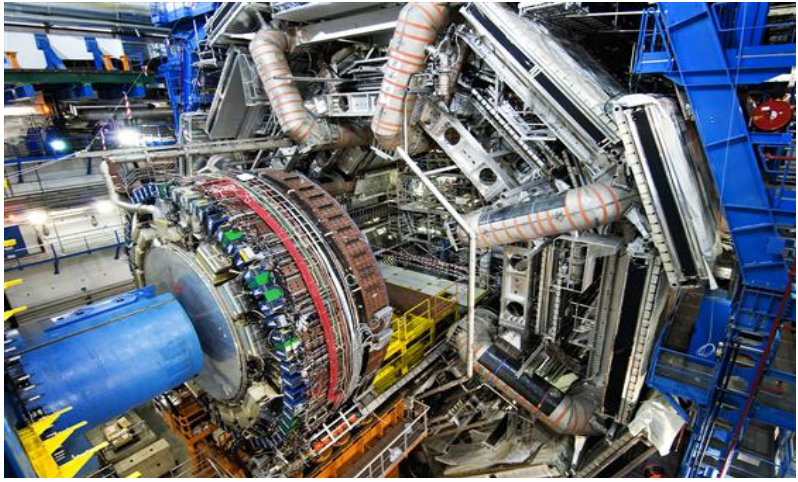

**Fig. 17:** Particle detector of the ATLAS collaboration at the LHC storage ring

The lattice has to be modified before and after the interaction point, to establish a large drift space in which the detector of the high-energy experiment can be embedded. At the same time the beams have to be focused strongly to get very small beam dimensions in both transverse planes at the collision point or in other words to get high luminosity. Such a lattice structure is called 'mini beta insertion'.

The luminosity of a particle collider relates the event rate *R* of a special reaction (e.g. a particle that is produced in the collision of the beams) and the cross-section $\Sigma_{\text{react}}$ of that physics reaction:

$$R = L * \Sigma_{\text{react}}$$

and is the result of the lattice design: the luminosity *L* of the storage ring. It is given by the beam optics at the collision point and the amount of the stored beam currents [9]:

$$L = \frac{1}{4\pi e^2 f_0 n_b} * \frac{I_{p1} I_{p2}}{\sigma^*_x \sigma^*_y}.$$

Here $I_1$ and $I_2$ are the values of the stored beam currents, $f_0$ is the revolution frequency of the machine and *b* the number of stored bunches. The terms $\sigma_x^*$ and $\sigma_y^*$ in the denominator are the beam sizes in the horizontal and vertical plane at the interaction point. For a high-luminosity collider the stored beam currents have to be high and at the same time the beams have to be focused at the interaction point to very small values.

Figure 18 shows the typical layout of such a mini beta insertion. It consists in general of:
i) a symmetric drift space that is large enough to house the particle detector and whose beam waist $\alpha_0 = 0$ is centred at the interaction point of the colliding beams;
ii) a quadrupole doublet (or triplet) on each side as close as possible;
iii) additional quadrupole lenses to match the Twiss parameters of the mini beta insertion to the optical parameters of the lattice cell in the arc.

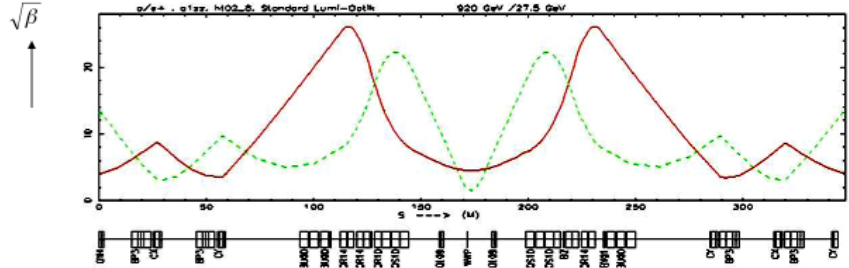

**Fig. 18**: Layout of a mini beta insertion scheme

As a mini beta scheme is always a kind of symmetric drift space we can apply the formula that we have derived above. For $\alpha = 0$ we get a quadratic increase of the $\beta$ function in the drift and at the distance $l_1$ of the first quadrupole lens we get

$$\beta(s) = \beta_0 + \frac{l_1^2}{\beta_0}.$$

The size of the beam at the position of the second quadrupole can be calculated in a similar way. According to Fig. 19 the transfer matrix of the quadrupole doublet system consists of four parts: two drifts with the lengths $l_1$ and $l_2$ and a focusing and defocusing quadrupole magnet. Starting at the IP we get, again in thin lens approximation,

$$M_{d1} = \begin{pmatrix} 1 & l_1 \\ 0 & 1 \end{pmatrix}, \qquad M_{q1} = \begin{pmatrix} 1 & 0 \\ 1/f_1 & 1 \end{pmatrix},$$

$$M_{d1} = \begin{pmatrix} 1 & l_2 \\ 0 & 1 \end{pmatrix}, \qquad M_{q2} = \begin{pmatrix} 1 & 0 \\ -1/f_2 & 1 \end{pmatrix}.$$

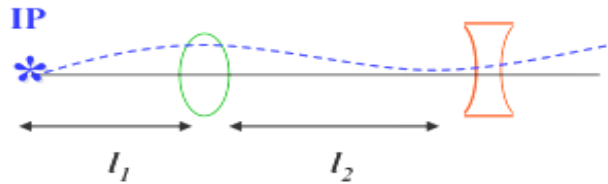

**Fig. 19:** Schematic layout of a mini beta doublet

It should be noted that in general the first lens of such a system is focusing in the vertical plane and therefore according to the sign convention used in this school the focal length is positive, $1/f_1 > 0$. The matrix for the complete system is

$$M = M_{QF} \cdot M_{D2} \cdot M_{QD} \cdot M_{D1}$$

$$M = \begin{pmatrix} 1 & 0 \\ -1/f_2 & 1 \end{pmatrix} \cdot \begin{pmatrix} 1 & l_2 \\ 0 & 1 \end{pmatrix} \cdot \begin{pmatrix} 1 & 0 \\ 1/f_1 & 1 \end{pmatrix} \cdot \begin{pmatrix} 1 & l_1 \\ 0 & 1 \end{pmatrix}.$$

Multiplying out we get

$$M = \begin{pmatrix} C & S \\ C' & S' \end{pmatrix} = \begin{pmatrix} 1 + \dfrac{l_2}{f_1} & l_1 + l_2 + \dfrac{l_1 l_2}{f_1} \\ \dfrac{1}{f_1} - \dfrac{1}{f_2} - \dfrac{l_2}{f_1 f_2} & -\dfrac{l_2}{f_2} - \dfrac{l_1 l_2}{f_1 f_2} - \dfrac{l_2}{f_2} + \dfrac{l_1}{f_1} + 1 \end{pmatrix}.$$

Remembering the transformation of the Twiss parameters in terms of matrix elements

$$\begin{pmatrix} \beta \\ \alpha \\ \gamma \end{pmatrix}_s = \begin{pmatrix} C^2 & -2SC & S^2 \\ -CC' & SC' + CS' & -SS' \\ C'^2 & -2S'C' & S'^2 \end{pmatrix} \cdot \begin{pmatrix} \beta \\ \alpha \\ \gamma \end{pmatrix}_0$$

we obtain

$$\beta(s) = C^2 \beta_0 - 2SC \alpha_0 + S^2 \gamma_0.$$

Here the index '0' denotes the interaction point and '$s$' refers to the position of the second quadrupole lens. As we are starting at the IP where $\alpha_0 = 0$ and $\gamma_0 = 1/\beta_0$ we can simplify and get

$$\beta(s) = C^2 \beta_0 + S^2 / \beta_0,$$

$$\beta(s) = \beta_0 \cdot \left(1 + \frac{l_2}{l_1}\right)^2 + \frac{1}{\beta_0}\left(l_1 + l_2 + \frac{l_1 l_2}{f_1}\right)^2.$$

This formula for $\beta$ at the second quadrupole lens is very useful when the gradient and aperture of the mini beta quadrupole magnet have to be designed.

### 5.2.1  *Phase advance in a mini beta insertion*

Unlike the situation in the arc where the phase advance is a function of the focusing properties of the cell, in a mini beta insertion or in any long drift space it is quasi a constant: As we know from linear beam optics, the phase advance is given by

$$\varphi(s) = \int \frac{\mathrm{d}s}{\beta(s)}$$

and inserting $\beta(s)$ from Eq. (27), we get

$$\varphi(s) = \frac{1}{\beta_0} \int_0^{l_1} \frac{1}{1 + \dfrac{s^2}{\beta_0^2}} \mathrm{d}s,$$

$$\varphi(s) = \arctan \frac{l_1}{\beta_0}$$

where $l_1$ denotes the distance of the first focusing element from the IP, i.e. the length of the first drift space. In Fig. 20 we have plotted the phase advance as a function of $l$ for a $\beta$ function of 10 cm. If the length of the drift is large compared with the value of $\beta$ at the IP, which is usually the case, the phase advance is approximately 90° on each side. In other words, the tune of the accelerator will increase by half an integer within the complete drift space.

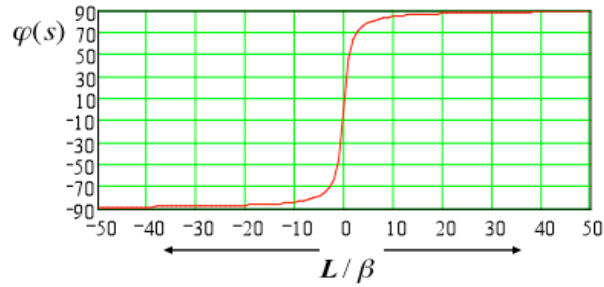

**Fig. 20:** Phase advance in a symmetric drift space as a function of the drift length

There are some more remarks concerning mini beta sections that will not be discussed here in detail but only mentioned briefly. As we have seen, large values of the *β* function on either side of the interaction point cannot be avoided if a mini beta section is inserted in the machine lattice. These high *β* values have a strong impact on the machine performance as detailed below.

i) According to Eq. (23) the chromaticity of a lattice is given by the strength of the focusing elements and the value of the *β* function at that position. In a mini beta insertion unfortunately we have both strong quadrupoles and large beam dimensions

$$Q' = -\frac{1}{4\pi} \oint \beta(s) k(s) \, ds.$$

The contribution of such a lattice section to $Q'$ can therefore be very large and as it has to be corrected in the ring it is in general one of the strong limitations of the luminosity in a collider ring.

ii) As in the insertion the beam dimension can reach large values, the aperture of the doublet magnets has to be much larger than in the FODO structure of the arc. Large magnet apertures however limit the strength of the quadrupole. Here a compromise has to be found between aperture requirement, integrated focusing strength, spot size at the IP, and financial cost, as large magnets are quite expensive.

iii) Last but not least I have to mention the problem of field quality and adjustment. Compared with the standard magnets in the arc, the lenses in a mini beta section have to fulfil stronger requirements. A kick due to a dipole error or which is adequate due to an off-centre quadrupole lens leads to an orbit distortion that is proportional to the *β* function at the place of the error (Eq. (22)).

The field quality concerning higher multipole components has to be much higher and the adjustment of the mini beta quadrupoles much more precise than those of quadrupole lenses in the arc. In general multipole components of the order of $\Delta B/B = 10^{-4}$ with respect to the main field and alignment tolerances in the transverse plane of about a tenth of a millimetre are desired.

## 6 Dispersion suppressors

The dispersion function *D*(*s*) has already been mentioned in Section 4.4 where we have shown that it is a function of the focusing properties of the lattice cell, and where we have calculated its size as a function of the cell parameters.

Now we have to come back to this topic in the context of lattice insertions: in the interaction region of an accelerator, that means in the straight section of a ring where two counter-rotating beams collide (typically designed as a mini beta insertion) the dispersion function *D*(*s*) has to disappear. A non-vanishing dispersion dilutes the luminosity of the machine and leads to additional stop bands in the working diagram of the accelerator (synchro-betatron resonances), that are driven by the beam–beam interaction.

Therefore sections have to be inserted in our magnet lattice that are designed to reduce the function $D(s)$ to zero, so-called dispersion suppressing schemes. In Eq. (19) we have shown that the oscillation amplitude of a particle is given by

$$x(s) = x_\beta(s) + D(s)\frac{\Delta p}{p_0}. \tag{29}$$

Here $x_\beta$ describes the solution of the homogeneous differential equation (which is valid for particles with ideal momentum $p_0$) and the second term, the dispersion term, describes the additional oscillation amplitude for particles with a relative momentum error $\Delta p/p_0$.

As an example let me present some numbers from the LHC proton storage ring: the beam size at the collision point of the two beams is in the horizontal and vertical direction given by the mini beta insertion: $\sigma_x = \sigma_y \approx 17$ µm. The contribution of the dispersion function to the particle amplitude with a typical dispersion in the cell of $D(s) \approx 1.5$ m and a momentum distribution of the beam $\Delta p/p \approx 5 \times 10^{-4}$ amounts to $x_D = 0.75$ mm. If not corrected the dispersion effect will dominate the beam size at the IP and lead to a considerable reduction of the achievable luminosity.

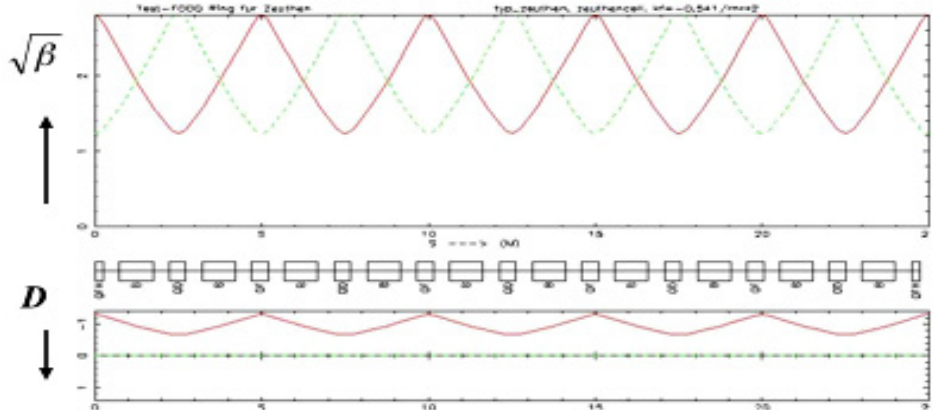

**Fig. 21:** Periodic FODO including the horizontal dispersion function in the lower part of the plot

## 6.1 Dispersion suppression using additional quadrupole magnets: 'the straightforward way'

There are several ways to suppress the dispersion and each of them has advantages and disadvantages. We will not present all of them but instead restrict ourselves to the basic idea behind it. Let us assume here that a periodic lattice is given and that one simply wants to continue this FODO structure of the arc through the straight section, but with vanishing dispersion.

Given an optical solution in the arc cells as for example shown in Fig. 21, we have to guarantee that starting from the periodic solution of the optical parameters $\alpha(s)$, $\beta(s)$, and $D(s)$ we obtain a situation at the end of the suppressor where we get $D(s) = D'(s) = 0$ and the values for $\alpha$ and $\beta$ unchanged.

The boundary conditions

$$D(s) = D'(s) = 0,$$
$$\beta_x(s) = \beta_{x\,\text{arc}}, \quad \alpha_x(s) = \alpha_{x\,\text{arc}},$$
$$\beta_y(s) = \beta_{y\,\text{arc}}, \quad \alpha_y(s) = \alpha_{y\,\text{arc}},$$

can be fulfilled by introducing six additional quadrupole lenses whose strengths have to be matched individually in an adequate way. This can be done by using one of the beam optics codes that are available today in every accelerator laboratory. An example is shown in Fig. 22, starting from a FODO structure with a phase advance of $\varphi \approx 61°$ per cell.

The advantages of this scheme are:

i) it works for arbitrary phase advance of the arc structure;

ii) matching works also for different optical parameters *α* and *β* before and after the dispersion suppressor;

iii) the ring geometry is unchanged as no additional dipoles are needed.

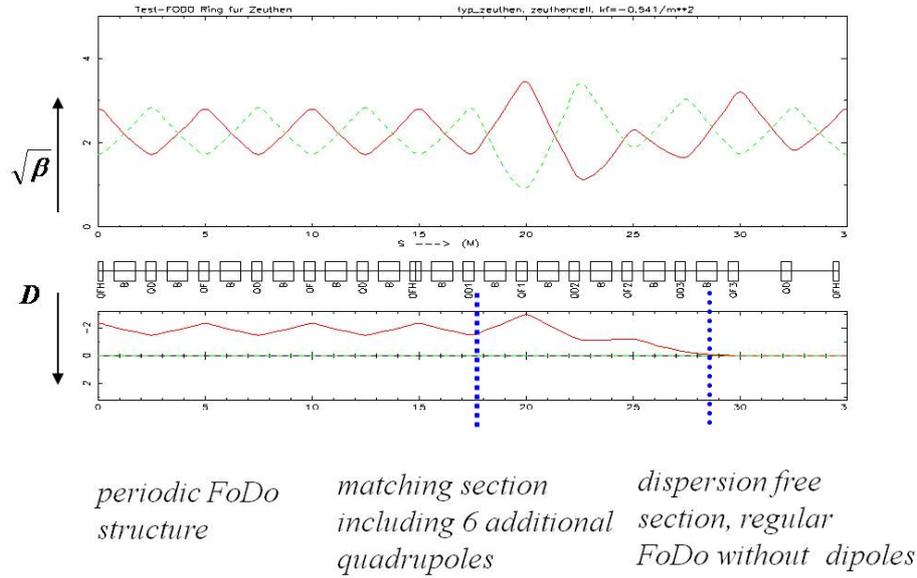

periodic FoDo structure | matching section including 6 additional quadrupoles | dispersion free section, regular FoDo without dipoles

**Fig. 22:** Dispersion suppressor using individual powered quadrupole lenses

On the other hand, there are a number of disadvantages that have to be mentioned:

i) as the strength of the additional quadrupole magnets have to be matched individually the scheme needs additional power supplies and quadrupole magnet types which can be an expensive requirement;

ii) the required quadrupole fields are in general stronger than in the arc;

iii) the *β* function reaches higher values (sometimes *really* high values) and so the aperture of the vacuum chamber and of the magnets has to be increased.

There are alternative ways to suppress the dispersion which do not need individually powered quadrupole lenses but instead change the strength of the dipole magnets at the end of the arc structure.

**6.2 'The clever way': half bend schemes**

This dispersion suppressing scheme exits of *n* additional FODO cells that are added to the periodic arc structure but where the bending strength of the dipole magnets is reduced. As before, we split the lattice into three parts: the periodic structure of the FODO cells in the arc, the lattice insertion where the dispersion is suppressed, followed by a dispersion-free section which can be another FODO structure without bending magnets or a mini beta insertion, etc.

The calculation of the suppressor goes in several steps.

*1) Establish the matrix for a periodic arc cell*

We have already calculated the dispersion in a FODO lattice before, see Eq. (21). In the thin lens approximation we have derived a formula for *D* as a function of the focusing properties of the lattice. Now we have to be a little bit more accurate and instead of the focusing strength and phase advance we have to work with the optical parameters of the system. We know that the transfer matrix in a lattice of a storage ring can be written as a function of the optical parameters Eq. (15)

$$M_{0 \to s} = \begin{pmatrix} \dfrac{\sqrt{\beta_s}}{\sqrt{\beta_0}} \cos\varphi + \alpha_0 \sin\varphi & \sqrt{\beta_0 \beta_s} \sin\varphi \\ \dfrac{(\alpha_0 - \alpha_s)\cos\varphi - (1+\alpha_0\alpha_s)\sin\varphi}{\sqrt{\beta_0 \beta_s}} & \dfrac{\sqrt{\beta_0}}{\sqrt{\beta_s}} \cos\varphi - \alpha_s \sin\varphi \end{pmatrix}. \tag{30}$$

The variable $\varphi$ refers to the phase advance between the starting point '0' and the end point '$s$' of the transformation. The formula is valid for any starting and end point in the lattice. If, for convenience, we refer the transformation to the middle of a focusing quadrupole magnet (as we usually did in the past) where $\alpha = 0$, and if we are interested in the solution for a complete cell, we can write the equation in a simpler form. Extending the matrix to the 3 × 3 form to include the dispersion terms and taking into account the periodicity of the system, $\beta_0 = \beta_s$ we get

$$M_{0 \to s} = \begin{pmatrix} C & S & D \\ C' & S' & D' \\ 0 & 0 & 1 \end{pmatrix} = \begin{pmatrix} \cos\phi_c & \beta_c \sin\phi_c & D(l) \\ \dfrac{-1}{\beta_c}\sin\phi_c & \cos\phi_c & D'(l) \\ 0 & 0 & 1 \end{pmatrix}. \tag{31}$$

Now $\Phi_C$ is the phase advance for a single cell and the index '$c$' reminds us that we talk about the periodic solution (one complete *cell*).

The dispersion elements $D$ and $D'$ are as usual given by the $C$ and $S$ elements according to Eq. (20):

$$D(l) = S(l) \cdot \int_0^l \frac{1}{\rho(\tilde{s})} C(\tilde{s})\, d\tilde{s} - C(l) \cdot \int_0^l \frac{1}{\rho(\tilde{s})} S(\tilde{s})\, d\tilde{s},$$

$$D'(l) = S'(l) \cdot \int_0^l \frac{1}{\rho(\tilde{s})} C(\tilde{s})\, d\tilde{s} - C'(l) \cdot \int_0^l \frac{1}{\rho(\tilde{s})} S(\tilde{s})\, d\tilde{s}.$$

The values $C(l)$ and $S(l)$ refer to the symmetry point of the cell (the middle of the quadrupole). The integral however has to be taken over the dipole magnet, where $\rho \neq 0$. Assuming a constant bending radius in the dipole magnets of the arc, $\rho$ = const (which is a good approximation in general), we can solve the integral over $C(s)$ and $S(s)$ if we approximate their values by those in the middle of the dipole magnet.

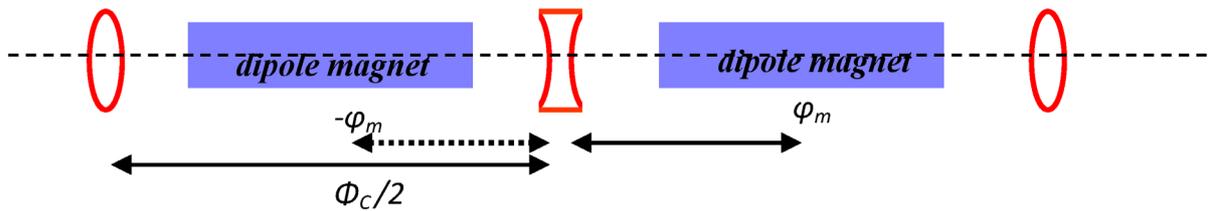

**Fig. 23:** Schematic view of a FODO: notation of the phase relations in the cell

*2) Transformation of the optical functions from the centre of the quadrupole to the middle of the dipole, to calculate the $C(\tilde{s})$ and $S(\tilde{s})$ functions*

As indicated in the schematic layout of Fig. 23 we have to transform the optical functions $\alpha$ and $\beta$ from the centre of the quadrupole lens to the centre of the dipole magnet. The formalism is given by Eq. (31) and we get (with $\alpha_0 = 0$)

$$C_m = \sqrt{\frac{\beta_m}{\beta_c}} \cos \Delta\phi = \sqrt{\frac{\beta_m}{\beta_c}} \cos\left(\frac{\phi_c}{2} \pm \varphi_m\right),$$

$$S_m = \sqrt{\beta_m \beta_c} \sin\left(\frac{\phi_c}{2} \pm \varphi_m\right).$$

The index *m* tells us that we are dealing with values in the middle of the bending magnets now and as our starting point was the centre of the QF quadrupole, the phase advance for this transformation is half the phase advance of the cell which brings us to the QD lens, plus/minus the phase distance $\varphi_m$ from that point to the dipole centre.

Now we can solve the integrals for $D(s)$ and $D'(s)$

$$D(l) = \beta_c \sin\phi_c \cdot \frac{L_B}{\rho} \sqrt{\frac{\beta_m}{\beta_c}} \cos\left(\frac{\phi_c}{2} \pm \varphi_m\right) - \cos\phi_c \cdot \frac{L_B}{\rho} \sqrt{\beta_m \beta_c} \sin\left(\frac{\phi_c}{2} \pm \varphi_m\right).$$
(32)

Here $L_B$ is the length of the dipole magnets and putting for the bending angle $\delta = L_B/\rho$ we get

$$D(l) = \delta \cdot \sqrt{\beta_m \beta_c} \left\{ \sin\phi_c \left[\cos\left(\frac{\phi_c}{2} + \varphi_m\right) + \cos\left(\frac{\phi_c}{2} - \varphi_m\right)\right] \right.$$

$$\left. - \cos\phi_c \left[\sin\left(\frac{\phi_c}{2} + \varphi_m\right) + \sin\left(\frac{\phi_c}{2} - \varphi_m\right)\right] \right\}.$$

Using the trigonometric relations

$$\cos x + \cos y = 2\cos\frac{x+y}{2} \cdot \cos\frac{x-y}{2}$$

$$\sin x + \sin y = 2\sin\frac{x+y}{2} \cdot \cos\frac{x-y}{2}$$

we get

$$D(l) = \delta \cdot \sqrt{\beta_m \beta_c} \left\{ \sin\phi_c \cdot 2\cos\frac{\phi_c}{2} \cos\varphi_m - \cos\phi_c \cdot 2\sin\frac{\phi_c}{2} \cos\varphi_m \right\}$$

$$D(l) = 2\delta \cdot \sqrt{\beta_m \beta_c} \cdot \cos\varphi_m \left\{ \sin\phi_c \cdot \cos\frac{\phi_c}{2} - \cos\phi_c \cdot \sin\frac{\phi_c}{2} \right\}$$

and with

$$\sin 2x = 2\sin x \cdot \cos x$$
$$\cos 2x = \cos^2 x - \sin^2 x$$

we can derive the dispersion at the middle of the quadrupole magnet in its final form

$$D(l) = 2\delta \cdot \sqrt{\beta_m \beta_c} \cdot \cos\varphi_m \left\{ 2\sin\frac{\phi_c}{2} \cdot \cos^2\frac{\phi_c}{2} - \left(\cos^2\frac{\phi_c}{2} - \sin^2\frac{\phi_c}{2}\right)\sin\frac{\phi_c}{2} \right\}$$

$$D(l) = 2\delta \cdot \sqrt{\beta_m \beta_c} \cdot \cos\varphi_m \sin\frac{\phi_c}{2} \left\{ 2\cos^2\frac{\phi_c}{2} - \cos^2\frac{\phi_c}{2} + \sin^2\frac{\phi_c}{2} \right\}$$

$$D(l) = 2\delta \cdot \sqrt{\beta_m \beta_c} \cdot \cos\varphi_m \sin\frac{\phi_c}{2}. \tag{33}$$

This is the expression for the dispersion term of the matrix (31) at the centre of the quadrupole magnet, determined from the dipole strength $1/\rho$ and matrix elements C and S at the position of the dipole.

In full analogy one derives the formula for the derivative of the dispersion, $D'(s)$

$$D'(l) = 2\delta \cdot \sqrt{\beta_m / \beta_c} \cdot \cos\varphi_m \cos\frac{\phi_c}{2}. \tag{34}$$

As we refer to the situation in the middle of a quadrupole, the expressions for $D(s)$ and $D'(s)$ are valid for a periodic structure, namely one FODO cell. Therefore, we require periodic boundary conditions for the transformation from one cell to the next:

$$\begin{pmatrix} D_c \\ D'_c \\ 1 \end{pmatrix} = M_c \cdot \begin{pmatrix} D_c \\ D'_c \\ 1 \end{pmatrix}$$

and by symmetry

$$D'_c = 0. \tag{35}$$

With these boundary conditions the periodic dispersion in the FODO cell is determined:

$$D_c = D_c \cdot \cos\phi_c + \delta \cdot \sqrt{\beta_m \beta_c} \cdot \cos\varphi_m 2\sin\frac{\phi_c}{2},$$

$$D_c = \delta \cdot \sqrt{\beta_m \beta_c} \cdot \cos\varphi_m / \sin\frac{\phi_c}{2}. \tag{36}$$

*3) Calculate the dispersion in the suppressor part*

In the dispersion suppressor section, $D(s)$ starting with the value at the end of the cell is reduced to zero. Or turning it around and thinking from right to left: the dispersion has to be created, starting from $D = D' = 0$. The goal will be to generate the dispersion in this section in a way that the values of the periodic arc cell are obtained.

The relation for $D(s)$ still holds in the same way

$$D(l) = S(l) \cdot \int_0^l \frac{1}{\rho(\tilde{s})} C(\tilde{s}) \, d\tilde{s} - C(l) \cdot \int_0^l \frac{1}{\rho(\tilde{s})} S(\tilde{s}) \, d\tilde{s},$$

but now we can take several cells into account (the number of cells inside the suppressor scheme) and we will have the freedom to choose a dipole strength $\rho_{\text{supr}}$ in this section which differs from the strength of the arc dipoles. As the dispersion is generated in a number of $n$ cells the matrix for these $n$ cells is

$$M_n = M_c^n = \begin{pmatrix} \cos n\phi_c & \beta_c \sin n\phi_c & D_n \\ \dfrac{-1}{\beta_c} \sin n\phi_c & \cos n\phi_c & D'_n \\ 0 & 0 & 1 \end{pmatrix}$$

and according to (32) the dispersion created in these $n$ cells is given by

$$D_n = \beta_c \sin n\phi_c \cdot \delta_{\text{suppr}} \cdot \sum_{i=1}^{n} \sqrt{\dfrac{\beta_m}{\beta_c}} \cos\left( i\phi_c - \dfrac{1}{2}\phi_c \pm \varphi_m \right)$$

$$- \cos n\phi_c \cdot \delta_{\text{suppr}} \cdot \sum_{i=1}^{n} \sqrt{\beta_m \beta_c} \sin\left( i\phi_c - \dfrac{1}{2}\phi_c \pm \varphi_m \right),$$

$$D_n = \sqrt{\beta_m \beta_c} \sin n\phi_c \cdot \delta_{\text{suppr}} \cdot \sum_{i=1}^{n} \cos\left( (2i-1)\dfrac{\phi_c}{2} \pm \varphi_m \right)$$

$$- \sqrt{\beta_m \beta_c} \cos n\phi_c \cdot \delta_{\text{suppr}} \cdot \sum_{i=1}^{n} \sin\left( (2i-1)\dfrac{\phi_c}{2} \pm \varphi_m \right).$$

Remembering the trigonometric gymnastics shown above we get

$$D_n = \sqrt{\beta_m \beta_c} \sin n\phi_c \cdot \delta_{\text{suppr}} \cdot \sum_{i=1}^{n} \cos\left( (2i-1)\dfrac{\phi_c}{2} \right) \cdot 2\cos\varphi_m$$

$$- \sqrt{\beta_m \beta_c} \cos n\phi_c \cdot \delta_{\text{suppr}} \cdot \sum_{i=1}^{n} \sin\left( (2i-1)\dfrac{\phi_c}{2} \right) \cdot 2\cos\varphi_m,$$

$$D_n = 2\sqrt{\beta_m \beta_c} \cdot \delta_{\text{suppr}} \cdot \cos\varphi_m \left\{ \sum_{i=1}^{n} \cos\left( (2i-1)\dfrac{\phi_c}{2} \right) \sin n\phi_c - \sum_{i=1}^{n} \sin\left( (2i-1)\dfrac{\phi_c}{2} \right) \cdot \cos n\phi_c \right\},$$

$$D_n = 2\sqrt{\beta_m \beta_c} \cdot \delta_{\text{suppr}} \cdot \cos\varphi_m \sin n\phi_c \dfrac{\sin\dfrac{n\phi_c}{2}\cos\dfrac{n\phi_c}{2}}{\sin\dfrac{\phi_c}{2}}$$

$$- 2\sqrt{\beta_m \beta_c} \cdot \delta_{\text{suppr}} \cdot \cos\varphi_m \cos n\phi_c \dfrac{\sin\dfrac{n\phi_c}{2}\sin\dfrac{n\phi_c}{2}}{\sin\dfrac{\phi_c}{2}},$$

$$D_n = \dfrac{2\sqrt{\beta_m \beta_c} \cdot \delta_{\text{suppr}} \cdot \cos\varphi_m}{\sin\dfrac{\phi_c}{2}} \left\{ 2\sin\dfrac{n\phi_c}{2}\cos\dfrac{n\phi_c}{2} \cdot \cos\dfrac{n\phi_c}{2}\sin\dfrac{n\phi_c}{2} \right.$$

$$\left. - \left( \cos^2\dfrac{n\phi_c}{2} - \sin^2\dfrac{n\phi_c}{2} \right) \sin^2\dfrac{n\phi_c}{2} \right\}.$$

And, finally,

$$D_n = \frac{2\sqrt{\beta_m \beta_c} \cdot \delta_{\text{suppr}} \cdot \cos\varphi_m}{\sin\frac{\phi_c}{2}} \sin^2 \frac{n\phi_c}{2}. \tag{37}$$

This relation gives us the dispersion $D(s)$ that is created in a number of $n$ cells that have a phase advance of $\Phi_C$ per cell. Here $\delta_{\text{supr}}$ is the bending strength of the dipole magnets located in these $n$ cells and the optical functions $\beta_m$ and $\beta_C$ refer to the values at the centre of the dipole and the quadrupole, respectively.

In a similar calculation we obtain the expression for the derivative $D'(s)$ of the dispersion:

$$D'_n = \frac{2\sqrt{\beta_m / \beta_c} \cdot \delta_{\text{suppr}} \cdot \cos\varphi_m}{\sin\frac{\phi_c}{2}} \sin n\phi_c. \tag{38}$$

*4) Determine the strength of the suppressor dipoles*

The last step is to calculate the strength of the dipole magnets in the suppressor section. As for the optimum match of $D$ the dispersion generated in this section has to be equal to that of the arc cells we equate the expressions (35), (36) and (37), (38) and get for $D_n$ the condition

$$D_n = \frac{2\sqrt{\beta_m \beta_c} \cdot \delta_{\text{suppr}} \cdot \cos\varphi_m}{\sin\frac{\phi_c}{2}} \sin^2 \frac{n\phi_c}{2} = \delta_{\text{arc}} \sqrt{\beta_m \beta_c} \frac{\cos\varphi_m}{\sin\frac{\phi_c}{2}}$$

and for $D'$

$$D'_n = \frac{2\sqrt{\beta_m / \beta_c} \cdot \delta_{\text{suppr}} \cdot \cos\varphi_m}{\sin\frac{\phi_c}{2}} \sin n\phi_c = 0.$$

From these last equations we deduce two conditions for the dispersion matching

$$\left. \begin{array}{l} 2\delta_{\text{suppr}} \sin^2\left(\dfrac{n\phi_c}{2}\right) = \delta_{\text{arc}} \\ \sin(n\phi_c) = 0 \end{array} \right\} \quad \rightarrow \quad \delta_{\text{suppr}} = \frac{1}{2}\delta_{\text{arc}}. \tag{39}$$

If the phase advance per cell in the arc fulfils the condition $\sin(n\Phi_C) = 0$, the strength of the dipoles in the suppressor region is just half the strength of the arc dipoles. In other words, the phase has to fulfil the condition

$$n\phi_c = k \cdot \pi, \quad k = 1, 3, \ldots .$$

There are a number of possible phase advances that fulfil that relation, but clearly not every arbitrary phase is allowed. Possible constellations would be, for example, $\Phi_C = 90°$, $n = 2$ cells or $\Phi_C = 60°$, $n = 3$ cells in the suppressor.

Figure 24 shows such a half bend dispersion suppressor, starting from a FODO structure with 60° phase advance per cell. The focusing strength of the FODO cells before and after the suppressor are identical, with the exception that, clearly, the FODO cells on the right are 'empty', i.e. they have no bending magnets.

It is evident that unlike in the suppressor scheme with quadrupole lenses now the $\beta$ function is unchanged in the suppressor region.

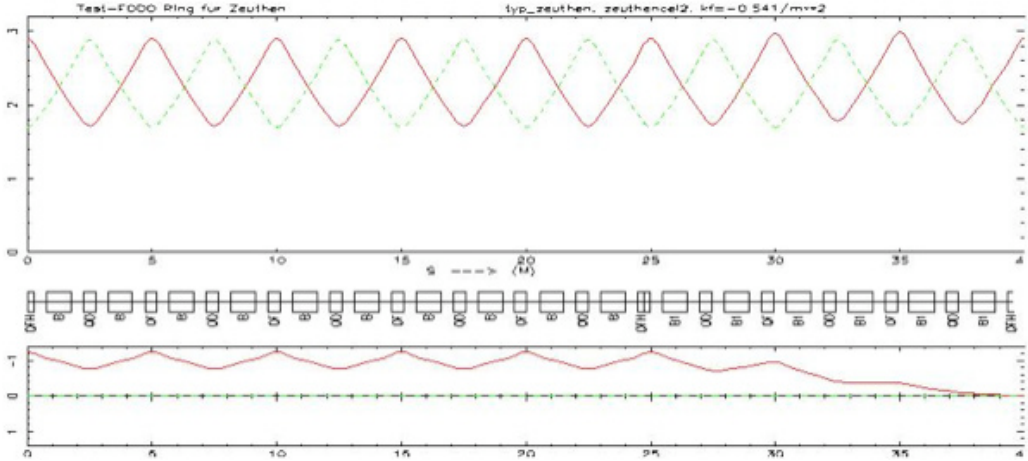

**Fig. 24:** Dispersion suppressor based on the half bend scheme

Again this scheme has advantages:

i) no additional quadrupole lenses are needed and no individual power supplies;

ii) the aperture requirements are just the same as in the arc, as the $\beta$ functions are unchanged;

and disadvantages:

i) it works only for certain values of the phase advance in the structure and therefore restricts the free choice of the optics in the arc;

ii) special dipole magnets are needed (having half the strength of the arc types);

iii) the geometry of the ring is changed.

I want to mention here, for purists only, that in theses equations the phase advance of the suppressor part is equal to that of the arc structure, which is not completely true as the weak focusing term $1/\rho^2$ in the arc FODO differs from the term $1/(2\rho)^2$ in the half bend scheme. As however the impact of the weak focusing on the beam optics can be neglected in many practical cases, Eq. (38) is *nearly* correct.

The application of such a scheme is very elegant but as it has a strong impact on the beam optics and geometry it has to be embedded in the accelerator design at an early stage.

## 6.3 The missing bend dispersion suppressor scheme

For completeness I would like to present another suppressor scheme, which is also used in a number of storage rings. It consists of a number of *n* cells without dipole magnets at the end of the arc, followed by *m* cells that are identical to the arc cells. The matching condition for this '*missing bend scheme*' with respect to the phase advance is

$$\frac{2n+m}{2}\phi_c = (2k+1)\frac{\pi}{2}$$

and for the number *m* of the required cells

$$\sin\frac{m\phi_c}{2} = \frac{1}{2}, \quad k = 0, 2... \quad \text{or} \quad \sin\frac{m\phi_c}{2} = \frac{-1}{2}, \quad k = 1, 3....$$

An example that is based on $\Phi = 60°$ and $m = n = 1$ is shown in Fig. 25.

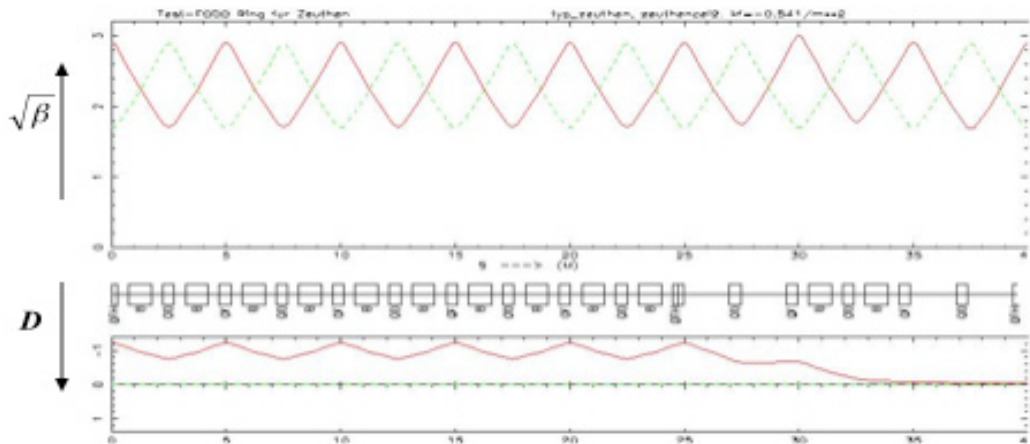

**Fig. 25:** Dispersion suppressor based on the missing magnet scheme

There are more scenarios for a variety of phase relations in the arc and the corresponding bending strength needed to reduce $D(s)$, see [10, 11].

In general, one will combine one of the two schemes (missing or half bend suppressor) with a certain number of individual quadrupole lenses to guarantee the flexibility of the system with respect to phases changes in the lattice and to keep the size of $\beta$ function moderate.